\documentclass[aps,prl,reprint,superscriptaddress,10pt]{revtex4-2}

\usepackage{graphicx,amsmath,amssymb,dcolumn,physics,color}

\DeclareMathOperator*{\argmax}{arg\,max}

\graphicspath{{./figs/}{./}}

\begin{document}

\title{
    Energy cost for flocking of active spins: the cusped dissipation maximum at the flocking transition
}

\author{Qiwei Yu}
\affiliation{IBM T.~J.~Watson Research Center, Yorktown Heights, NY 10598}
\affiliation{Lewis-Sigler Institute for Integrative Genomics, Princeton University, Princeton, NJ 08544}

\author{Yuhai Tu}
\affiliation{IBM T.~J.~Watson Research Center, Yorktown Heights, NY 10598}


\begin{abstract}
    We study the energy cost of flocking in the active Ising model (AIM) and show that besides the energy cost for self-propelled motion, an additional energy dissipation is required to power the alignment of spins.
    We find that this additional alignment dissipation reaches its maximum at the flocking transition point in the form of a cusp with a discontinuous first derivative with respect to the control parameter.
    To understand this singular behavior, we analytically solve the two- and three-site AIM models and obtain the exact dependence of the alignment dissipation on the flocking order parameter and control parameter, which explains the cusped dissipation maximum at the flocking transition. Our results reveal a trade-off between the energy cost of the system and its performance measured by the flocking speed and sensitivity to external perturbations. This trade-off relationship provides a new perspective for understanding the dynamics of natural flocks and designing optimal artificial flocking systems.
\end{abstract}

\maketitle
\clearpage

Understanding how collective coherent motion (``flocking'') emerges from a system of self-propelled, interacting individuals has been a central question in nonequilibrium statistical physics and biophysics~\cite{toner_hydrodynamics_2005,ramaswamy_mechanics_2010,marchetti_hydrodynamics_2013}.
Familiar examples include birds, fish, bacteria~\cite{marchetti_hydrodynamics_2013,bialek_statistical_2012}, and synthetic systems such as active colloids~\cite{kaiser_flocking_2017}. Theoretical studies have involved models of self-propelled, aligning particles with continuous~\cite{vicsek_novel_1995,toner_long-range_1995,toner_flocks_1998,gregoire_onset_2004,chate_collective_2008} or discrete~\cite{solon_revisiting_2013,solon_flocking_2015} symmetry.
Despite their diversity, these systems are all far from thermodynamic equilibrium~\cite{gnesotto_broken_2018} and thus a continuous dissipation of free energy is required to create and maintain the long-range flocking order.
Indeed, energy dissipation plays a crucial role in driving living systems out of equilibrium to achieve important biological functions, such as adaptation~\cite{lan_energyspeedaccuracy_2012}, error correction~\cite{hopfield_kinetic_1974,ninio_kinetic_1975,bennett_dissipation-error_1979,murugan_speed_2012,sartori_thermodynamics_2015,yu_energy_2022}, spatial patterns~\cite{falasco_information_2018}, and temporal oscillation~\cite{cao_free-energy_2015}. Here, we study the nonequilibrium thermodynamics of dry aligning active matter~\cite{chate_dry_2020} aiming to elucidate the relationship between the energetic cost of flocking and its performance measured by the flocking speed and sensitivity.

The {dynamics and energy dissipation (entropy production) of flocking} can be studied at the microscopic level by prescribing the single-particle dynamics~\cite{vicsek_novel_1995} or at the coarse-grained level with hydrodynamic field theories~\cite{toner_long-range_1995,toner_flocks_1998}.
{In the latter case, the entropy production rate (EPR) calculated from the standard procedure gives a measure of irreversibility but usually has no thermodynamic interpretation (unless under special conditions such as linear irreversible thermodynamics~\cite{markovich_thermodynamics_2021}).
Namely, it does not give the (physical) heat dissipation rate, and an alternative term ``information EPR'' has been proposed to differentiate it from the microscopic EPR, which has unambiguous thermodynamic interpretation~\cite{seifert_stochastic_2012,nardini_entropy_2017,fodor_irreversibility_2022}.
The reason behind this discrepancy is that coarse graining drastically decreases the dissipation rate~\cite{yu_inverse_2021,yu_state-space_2022,cocconi_scaling_2022}, which means that macroscopic theories tend to dramatically underestimate the energy dissipation.
Therefore, it is fundamentally important to elucidate the energy cost of flocking using a microscopic model, which gives the ``true'' heat dissipation, despite existing work using hydrodynamic approaches~\cite{fodor_how_2016,borthne_time-reversal_2020}.}

Here, we investigate the energy dissipation of the active Ising model (AIM)~\cite{solon_revisiting_2013,solon_flocking_2015} which describes a lattice gas of Ising spins with ferromagnetic alignment and biased diffusion. Our main finding is a cusped energy dissipation maximum at the flocking transition point, which is supported by numerical simulations and confirmed by the analytical solution of reduced AIMs with two or three sites. These findings uncover a new perspective on the energy-speed-sensitivity trade-off in flocking.

\textbf{Dissipation in the active Ising model.}
The 2D AIM describes $N$ particles on an $L_x\times L_y$ lattice with periodic boundary conditions. Each particle carries an Ising spin $s=\pm 1$, and the number of $\pm$ spins on site $(i,j)$ is denoted by $n_{i,j}^\pm$ (no volume exclusion). The system follows continuous-time Markovian dynamics including flipping (local alignment) and hopping (self-propulsion).
Each particle can flip its spin from $s$ to $(-s)$ at rate $\omega e^{-\beta E_0 s m_{i,j}/\rho_{i,j}}$, where $m_{i,j} = n_{i,j}^+-n_{i,j}^-$ and $\rho_{i,j} = n_{i,j}^++n_{i,j}^-$ are the local magnetization and density, respectively. $\omega^{-1}$ sets the flipping timescale. $E_0$ measures the strength of the spin-spin alignment interaction, and $\beta$ is the inverse temperature which is set to $1$.
Each spin can also hop to one of the four neighboring sites, at rate $D(1+s\epsilon)$ to the right, $D(1-s\epsilon)$ to the left, and $D$ to up and down.
The flipping dynamics obeys detailed balance according to the Hamiltonian of a fully connected (mean-field) Ising model, but the hopping dynamics breaks detailed balance and drives the system out of equilibrium.
Flocking is defined as the emergence of long-range order (LRO) among spins characterized by a finite $\expval{s}$ with the mean flocking speed given by $v=2D\epsilon \expval{s}$ in the $x$ direction. Note that in the special case of unbiased diffusion $\epsilon=0$, LRO can still emerge albeit with zero flocking speed [see Fig.~S3 in the Supplemental Material (SM) for details].

Two equivalent approaches are employed to calculate the steady-state entropy production (energy dissipation) rate.
The first method calculates the average dissipation rate from the ratio of forward and backward realizations of a sufficiently long trajectory (assuming ergodicity) obtained by simulating the AIM dynamics~\cite{peliti_stochastic_2021}.
The second approach considers the different spin configurations ($\qty{n^{\pm}_{i,j}}$) as states of a reaction network with flipping and hopping as the two types of transitions between different states. Once the AIM reaction network reaches its nonequilibrium steady state (NESS), the dissipation rate can be determined by following the standard procedure for computing entropy production rate of reaction networks~\cite{hill_free_1977,qian_open-system_2006}.
These two approaches are equivalent. The former is suited for the numerical simulation of the full AIM, and the latter offers analytical tractability in the two-site (and three-site) AIM.

\begin{figure}
    \includegraphics[width=\linewidth]{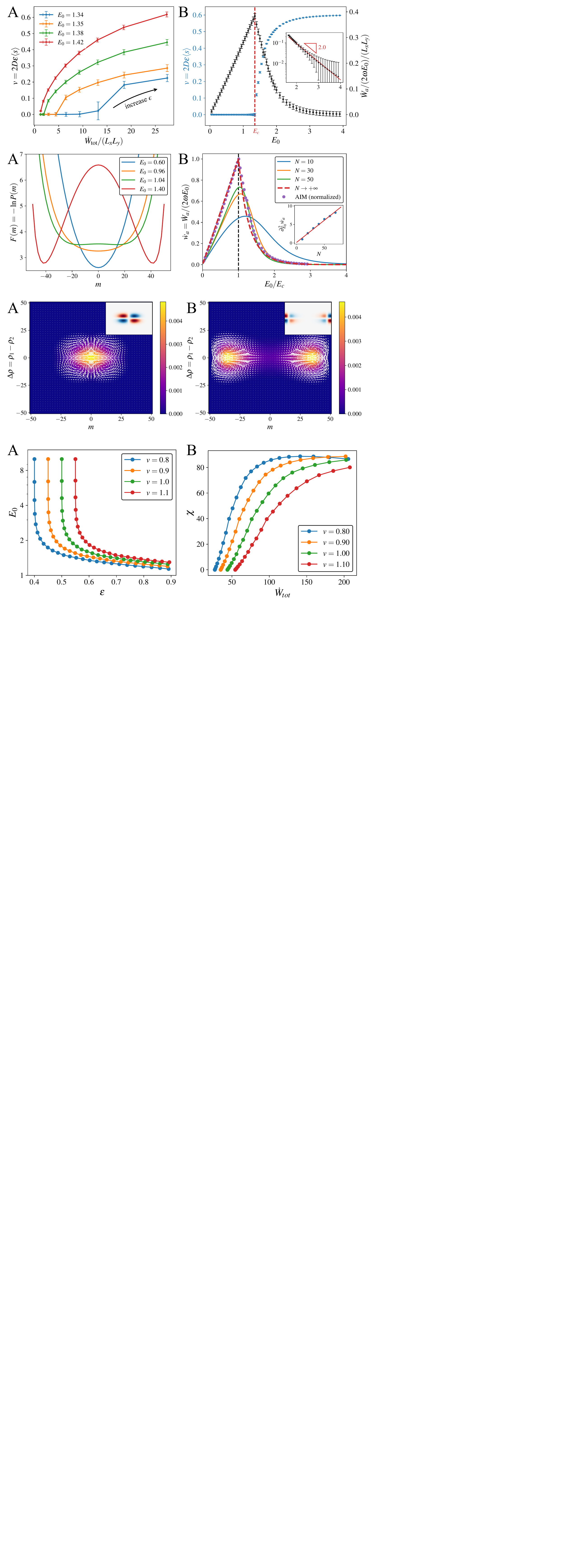}
    \caption{
        (A) The average flocking speed $v$ versus the total dissipation $\dot{W}_\mathrm{tot}$ for fixed values of $E_0$ and increasing $\epsilon$.
        (B) The average flocking speed (blue) and alignment dissipation (black) for $\epsilon=0.3$. The red dashed line is the transition point $E_c$ above which $v>0$. The inset shows the exponential decay of dissipation at large $E_0$.
        $L_x=300$, $L_y=100$, $\bar{\rho} = N/(L_xL_y)=5$, $D=1$, $\omega=1$. {See section IA in the Supplemental Material for details.}
    }
    \label{Fig1: full AIM dissipation}
\end{figure}

A finite amount of energy dissipation is needed to drive the system sufficiently away from equilibrium to generate flocking behavior.
As shown in Fig.~\ref{Fig1: full AIM dissipation}A, a nonzero flocking speed $v$ can be achieved by increasing $\epsilon$ at fixed $E_0$, which also increases the total dissipation rate $\dot{W}_\mathrm{tot}$.
The flocking motion does not emerge until $\dot{W}_\mathrm{tot}$ is above a certain (nonzero) threshold.

The total dissipation rate can be decomposed into contributions from the two types of transitions: $\dot{W}_\mathrm{tot}=\dot{W}_{m}+\dot{W}_a$, where $\dot{W}_m$ and $\dot{W}_{a}$ correspond to the dissipation rates due to motion (hopping) and alignment (flipping) of the particles, respectively.
Since each particle moves at an average speed $v_0=2D\epsilon$ and each step along the bias direction costs energy $\ln\frac{1+\epsilon}{1-\epsilon}$, the resulting dissipation rate for motion is simply $\dot{W}_{m}=Nv_0\ln\frac{1+\epsilon}{1-\epsilon}$ {(see Appendix A for details)}.
The alignment dissipation $\dot{W}_a$ can be calculated by summing up the cost of all flipping events during a sufficiently long time interval $\tau$:
\begin{align}
    \dot W_a
    = \lim\limits_{\tau\to\infty} \frac{1}{\tau} \sum_{0<t<\tau} 2E_0 \frac{1-m_{i,j}s}{\rho_{i,j}}.
    \label{Eq: wdot flip definition}
\end{align}
Each event flips a spin $s$ to $(-s)$ on site $(i,j)$, which has local magnetization $m_{i,j}$ and local density $\rho_{i,j}$ {(see Appendix A)}.
It will be convenient to henceforth refer to the nondimensionalized alignment dissipation rate $\dot{w}_a=\dot{W}_a/(2\omega E_0)$ as the alignment dissipation.

The motion dissipation rate $\dot{W}_m$ is responsible for driving the self-propulsion of the particles, which is independent of the alignment dynamics. As expected, $\dot{W}_m$ increases monotonically with $\epsilon$, vanishing in the unbiased limit ($\epsilon=0$) and diverging in the irreversible limit ($\epsilon\to 1$).
The origin of the alignment dissipation $\dot{W}_a$ is more subtle. Although the local spin flipping dynamics obeys detailed balance, the local spin system at a given site is driven out of equilibrium by the continuous exchange of spins between neighboring sites due to the transport process. A continuous dissipation rate $\dot{W}_a$ is needed to drive the spin alignment to maintain the flocking order. As a result, $\dot{W}_a$ depends on both the alignment strength ($E_0$) and the particle's key transport properties, in particular, the motion bias $\epsilon$ and the relative timescale $D/\omega$.
Next, we investigate how the flocking behavior and dissipation rates depend on the these key control parameters of the system ($E_0$, $\epsilon$, $D/\omega$).

\textbf{A cusped dissipation maximum at the flocking transition.}
For a fixed bias $\epsilon$, the system remains disordered ($v=0$) until $E_0$ is increased above a certain threshold $E_c$ (Fig.~\ref{Fig1: full AIM dissipation}B).
The alignment dissipation $\dot{w}_{a}$ increases linearly in the disordered phase and decreases monotonically in the flocking phase (exponentially at large $E_0$ as shown by the inset). Remarkably, it reaches maximum exactly at the transition point $E_c$ in the form of a cusp. The value of $\dot{w}_{a}$ is continuous across the transition, but its derivative $\dv{\dot{w}_{a}}{E_0}$ changes abruptly from positive to negative across $E_c$ forming a cusp at its maximum.
Since $\dot{W}_{m}$ stays constant, the same cusped maximum behavior also exists for $\dot{W}_\mathrm{tot}$.
Extensive numerical simulations find this behavior to be general, regardless of the bias $\epsilon$ or the relative timescale set by $D/\omega$ (see {Appendix A}).
The critical $E_c$ decreases with $\epsilon$ and increases with $\omega$, but it always coincides with the maximum of $\dot{w}_{a}$.
The alignment dissipation can be decomposed into the product of the frequency of flipping events $\dot{n}_f$ and the mean energy cost per flip $\bar{w}_f = \dot{w}_{a}/\dot{n}_f$.
At the transition point, they both have continuous values but discontinuous first derivatives, which results in the cusp of $\dot{w}_a$ (see the Supplemental Material).

As discussed previously, the key to understanding the alignment dissipation is how the transport of spins between neighboring sites drives the local spin system out of equilibrium. However, it is difficult to understand the full AIM with a large system size due to the numerous degrees of freedom. Next, we investigate the alignment dissipation in a reduced AIM with the minimal number of sites that allows transport of active spins.

\begin{figure}
    \includegraphics[width=\linewidth]{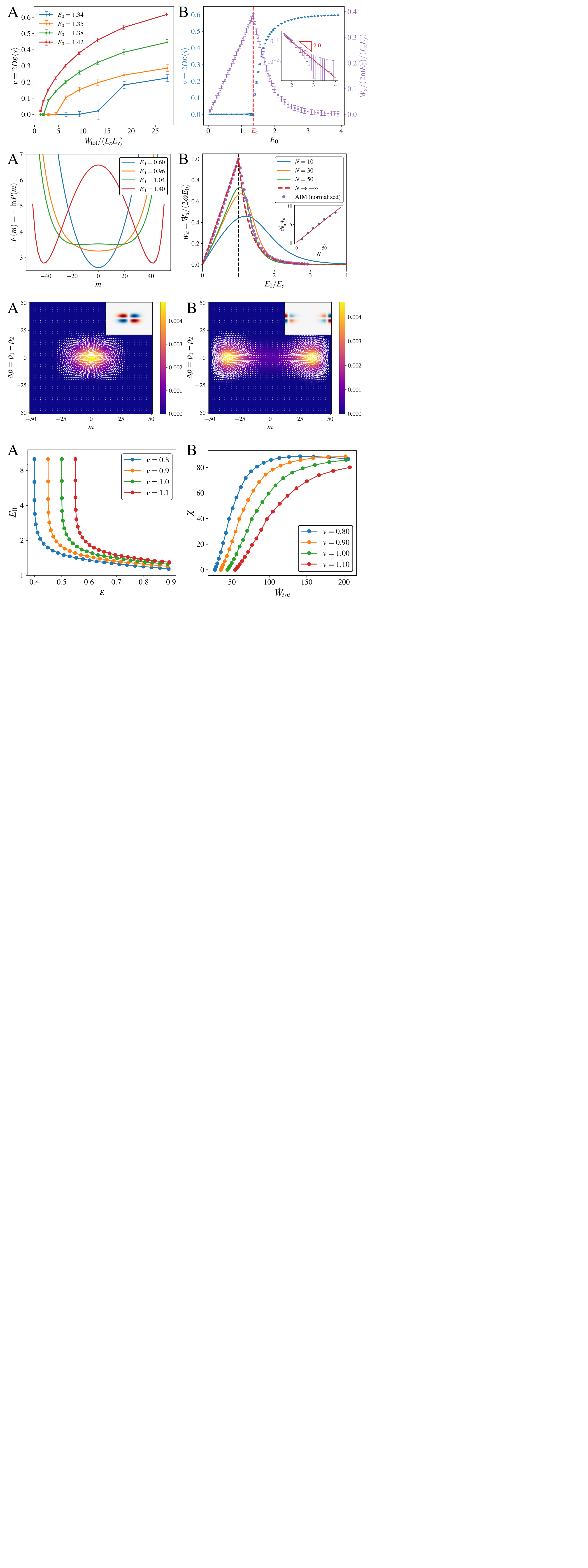}
    \caption{
        (A) The effective free energy landscape demonstrates the existence of a nonequilibrium phase transition in the two-site AIM.  $N=50$, $D=\omega=1$.
        (B) The alignment dissipation of the two-site AIM with different finite $N$ (solid lines) and infinite $N$ (red dashed line) and the full AIM (purple dots, normalized according to the text).
        Inset: the curvature at the dissipation maximum in the two-site AIM.
    }
    \label{Fig2: two-site}
\end{figure}

\textbf{The two-site (and three-site) AIM shows the cusped maximum of flocking dissipation.}
We consider a special case of the AIM with only two sites ($L_x=2$ and $L_y=1$), which is the minimum system size needed to drive the AIM out of equilibrium to produce flocking behavior. The flipping and hopping dynamics are the same as the full AIM. Importantly, hopping left- and rightward are considered to be two different processes even though they reach the same site due to the periodic boundary condition. In the flocking phase, the left-right symmetry is spontaneously broken and hopping in one direction dominates. Conceptually, the two-site AIM can be considered as a coarse-grained version of the full AIM. It retains much of the physics of the full AIM including the flocking transition and the associated dissipation maximum.

The model is fully characterized by the total number of spins $N$ and three state variables $(a_0,a_1,b_1)$ where $a_0$ is  the total number of $+$ spins; $a_1$ and $b_1$ correspond to the number of $+$ and $-$ spins on site 1, respectively. The dynamics of the probability distribution $P(a_0,a_1,b_1)$ is governed by the master equation,
\begin{equation}
    \dv{P(a_0,a_1,b_1)}{t} = \mathcal{L} P(a_0,a_1,b_1),
    \label{Eq:two-site masterEq L}
\end{equation}
where $\mathcal{L}$ is a linear operator (matrix) capturing the transitions. {(The full equation and its solution are covered in the Supplemental Material, Sec. II.)}
The steady-state distribution $P^s(a_0,a_1,b_1)$ can be found by solving $\mathcal{L} P^s(a_0,a_1,b_1)=0$ subject to normalization $\sum_{a_0,a_1,b_1} P^s(a_0,a_1,b_1)=1$ and can be used to compute all statistical properties of the system, e.g., the average total magnetization $\expval{m} =\sum_{a_0,a_1,b_1}(2a_0-N)P^s(a_0,a_1,b_1)$.

At finite $N$, the phase transition point $E_c$ can be determined by computing the effective free energy landscape $F(m)=-\ln P(m)$ where $P(m)=\sum_{a_0,a_1,b_1}\delta (2a_0-N-m)P^s(a_0,a_1,b_1)$ is the steady-state distribution of the total magnetization $m$. As shown in Fig.~\ref{Fig2: two-site}A, as $E_0$ increases, the disordered state $m=0$ goes from stable [$F''(0)>0$] to unstable [$F''(0)<0$], indicating the emergence of flocking.
The transition point $E_c$ (determined by $F''(0)=0$) and the position of the alignment dissipation maximum  ($E_m=\argmax_{E_0}\dot{w}_a$) are extrapolated to converge at infinite $N$ {(see the Supplemental Material, Fig.~S4)}.
Moreover, the curvature at the peak $\eval{\partial_{E_0}^2 \dot{w}_{a}}_{E_0=E_m}$ increases with $N$ (Fig.~\ref{Fig2: two-site}B inset) and it is projected to diverge at infinite $N$.
These results indicate a cusped dissipation maximum at flocking transition of the two-site AIM, consistent with observation in the full AIM.

In the infinite $N$ limit, the steady-state probability $P^s$ can be obtained analytically by assuming $D\gg \omega$. However, this assumption is not essential to the results. {Perturbation theory shows that higher-order corrections of the order $\mathcal{O}(\frac{\omega}{D})$ do not affect the cusped maximum behavior in $\dot{w}_a$(see Appendix B).}
The effective free energy in the limit $\omega/D\rightarrow 0$ is
\begin{equation}
    \frac{F(m)}{N} =z\ln{z}+(1-z)\ln\qty({1-z})+2E_0z(1-z)+O(N^{-1}),
    \label{Eq:free energy}
\end{equation}
where $z = a_0/N=(N+m)/(2N)$ is the fraction of spin-up.
The flocking transition takes place at $E_c=1$, where the most probable state (saddle point) goes from the disordered state ($z=\frac{1}{2}$) to the flocking state with $z=z^\star(\neq\frac{1}{2})$ where $z^\star$ is determined by
\begin{equation}
    \frac{1}{2(1-2z^\star)} \ln\frac{1-z^\star}{z^\star}= E_0, \quad (E_0>1),
\end{equation}
which has two solutions $z^\star$ and $(1-z^\star)$ corresponding to flocking left- and rightward, respectively.
Although the free energy is equivalent to that of the mean-field Ising model, the system continuously dissipates energy due to nonvanishing state-space fluxes. The fluxes associated with flipping give the alignment dissipation:
\begin{equation}
    \dot{w}_a = \frac{1}{2\omega E_0} \sum_{\mathrm{flip}} (J_+-J_-) \ln \frac{k_+}{k_-}= {\sum \sigma\cdot P^s \equiv \expval{\sigma}},
    \label{Eq:wa twosite definition}
\end{equation}
where $\sigma$ is the local alignment dissipation rate whose average over $P^s$ gives the steady-state alignment dissipation $\dot{w}_a$.
The averaging is computed using the saddle point method, which expands $\sigma$ around the most probable state.
Importantly, direct evaluation of $\sigma$ at the saddle point vanishes, and the leading-order contribution comes from expansion to the second order in $(a_1,b_1)$.
This indicates that particle number fluctuations are the essential source of alignment dissipation, which can be expressed as
\begin{widetext}
    \begin{equation}
        \begin{aligned}
            \dot{w}_a & = {\frac{1}{2}\qty[\pdv[2]{\sigma}{a_1}\expval{\qty(a_1-a_1^\star)^2}+ \pdv[2]{\sigma}{b_1}\expval{\qty( b_1-b_1^\star)^2}]
            = \begin{cases}
                E_0+\mathcal{O}(\frac{\omega}{D}),                                 & 0<E_0<E_c(=1) \\
                8E_0 \qty[z^\star(1-z^\star)]^{3/2}+\mathcal{O}(\frac{\omega}{D}), & E_0\ge E_c
            \end{cases},
            }
        \end{aligned}
        \label{cusp}
    \end{equation}
\end{widetext}
where the derivatives are evaluated at the saddle point $(a_0^\star,{a}_1^\star,b_1^\star)=(z^\star N, z^\star N/2, (1-z^\star)N/2)$. The explicit expressions for the $\mathcal{O}(\frac{\omega}{D})$ terms can be found in Appendix B and the Supplemental Material.

It is clear from Eq.~\ref{cusp} that $\partial _{E_0}\dot{w}_a$ is discontinuous at the critical point ($E_0=E_c$) because $\partial_{E_0} z^\star$ is discontinuous there. Quantitatively, we have $\partial _{E_0}\dot{w}_a|_{E_0=1^-}=1 $ and $\partial _{E_0}\dot{w}_a|_{E_0=1^+}=-3.5$, which shows that $\dot{w}_a$
(red dashed line in Fig.~\ref{Fig2: two-site}B) exhibits a cusped maximum exactly at $E_c=1$. Eq.~\ref{cusp} explicitly connects the dissipation $\dot{w}_a$ to number fluctuations ($\expval{(a_1-a_1^\star)^2}$ and $\expval{(b_1-b_1^\star)^2}$).

To make a direct comparison between the two-site AIM and the full AIM, we rescale $E_0$ by $E_c$, normalize the dissipation by its maximum, and plot them against each other in Fig.~\ref{Fig2: two-site}B.
The two models agree exactly in the disordered phase where dissipation grows linearly with $E_0$ as well as deep in the flocking phase where dissipation decays exponentially to zero. The cusped maximum at transition is also in good agreement, evident from the discontinuity of the slope.
There is a small quantitative difference in dissipation at $E_0$ slightly above $E_c$ because the two-site model cannot capture the flocking band structure in the mixed phase~\footnote{From the perspective of coarse-graining,
    the flocking phase of the two-site model includes both the mixed phase and the liquid phase of the full AIM.}.

Although the two-site AIM captures the flocking transition and the cusped dissipation maximum of the full AIM, $\dot{w}_a$  does not depend on the bias $\epsilon$ (Eq.~\ref{cusp}) since hopping to the left and to the right end up at the same site.  To make sure this special property of the two-site model does not affect the general results, we extend the analytical solution to the three-site AIM. Aside from being more tedious, the three-site AIM can be solved in a similar fashion as the two-site AIM (see Appendix B and the Supplemental Material for details), which not only confirms the existence of the cusped dissipation maximum at the flocking transition but also captures the dependence of $\dot{w}_a$ on $\epsilon$ explicitly.

\textbf{The energy-speed-sensitivity trade-off.}
The flocking of interacting particles is conceptually analogous to the synchronization of coupled oscillators~\cite{kuramoto_chemical_2003,zhang_energy_2020}, which can be understood as flocking (collective dynamics) of the phases of individual clocks.
In both cases, an extra energy dissipation is needed to maintain coherence among individual subsystems (spins or oscillators) that are already out of equilibrium. However, the dissipation of these two systems exhibits different behaviors. For coupled oscillators, the dissipation increases with the order parameter, meaning that it is very costly to maintain a system of highly coupled (and therefore synchronized) oscillators~\cite{zhang_energy_2020}. In the AIM, however, dissipation peaks exactly at the transition and decreases with interaction ($E_0$) in the flocking phase. At large $E_0$, the highly ordered flock requires a smaller energy to maintain.
The difference between the two behaviors stems from the alignment mechanisms. The active spins align locally, which effectively synchronizes their velocities.
The coupled oscillators are synchronized by exchanging phases, whose analogy in the AIM would be simultaneous displacement of pairs of particles. This nonlocal interaction couples the alignment cost to the cost of motion (i.e.~advancing the individual clocks), leading to a higher dissipation in the ordered phase.
These analyses suggest that compared to exchanging position, local alignment of velocity is an energetically more favorable way of maintaining the order in a system of active particles.

\begin{figure}
    \includegraphics[width=\linewidth]{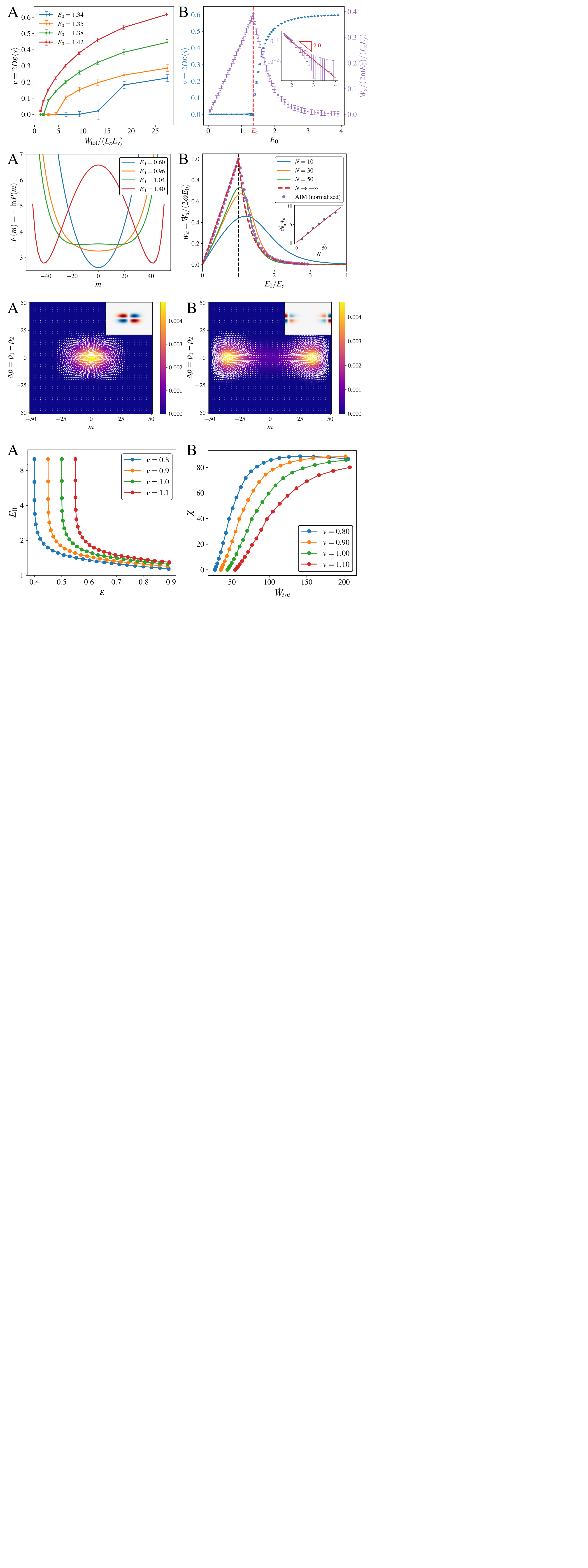}
    \caption{
        The energy-speed-sensitivity trade-off in the two-site AIM.
        (A) Contours for constant $v$ in the $(\epsilon, E_0)$ plane.
        (B) The total dissipation and sensitivity along different $v$ contours.
        $N=40$, $D=\omega=1$.
    }
    \label{Fig4: tradeoff}
\end{figure}

Another key property of flocks is its sensitivity to external perturbations, which we characterize by the magnetic susceptibility $\chi$ of the AIM.
In AIM, there are many choices of parameters $(\epsilon, E_0)$ to achieve any given flocking speed $v$ as shown in Fig.~\ref{Fig4: tradeoff}A. For a given $v$, the total dissipation achieves its minimum in the limit of $E_0\to\infty$ and $\epsilon\to v/(2D)$, which unfortunately leads to zero sensitivity ($\chi=0$).
However, sensitivity can be increased by decreasing $E_0$, which requires increasing $\epsilon$ in order to maintain a fixed $v$ (Fig.~\ref{Fig4: tradeoff}A).
As a result, $\dot{W}_m$ and thereby the total dissipation increases.
Fig.~\ref{Fig4: tradeoff}B demonstrates this trade-off whereby enhancing sensitivity at a constant flocking speed necessarily increases dissipation.
Similarly, for a given sensitivity, increasing the flocking speed also requires more dissipation; for a given dissipation, increasing speed necessarily reduces sensitivity (see the Supplemental Material, Sec. II for analytical expressions).
These relations constitute an energy-speed-sensitivity trade-off, which may affect the strategy for flocks (natural or artificial) to optimize their performance with limited resources.

\textbf{Discussion and future directions.}
A heuristic argument for the long-range order in the hydrodynamics theory of flocking (the Toner-Tu equation) was the stabilizing effect of the convective term, which enables particles to change their neighborhoods of interaction~\cite{toner_long-range_1995}.  Our study suggests that such a stabilizing mechanism that combines motion and alignment is intrinsically out of equilibrium and must be sustained by continuous energy dissipation ($\dot{w}_a$).
Specifically, the motion leads to number fluctuations which, as shown by Eq.~\ref{cusp}, directly causes energy dissipation. The fluctuation reaches maximum at the transition, resulting in the dissipation maximum. Therefore, the dissipation maximum reported here is deeply connected to the underlying mechanism that leads to flocking transition. Given that the same mechanism underlies general flocking models, it will be interesting to extend this study to flocking theories with continuous symmetry and off-lattice models~\cite{vicsek_novel_1995,toner_long-range_1995,toner_flocks_1998}. In fact, a recent study on the Vicsek model also finds dissipation maximum near flocking transition, which suggests that the phenomenon observed in this study may be general~\cite{ferretti_signatures_2022}.
Another possible direction is to compare the energy cost of flocking to other models of nonequilibrium phase transitions, some of which also demonstrate reduced energy dissipation in the ordered phase~\cite{herpich_collective_2018}.

The two-site (and three-site) AIM provides a useful approach for understanding spatially extended nonequilibrium systems without completely going to the mean-field limit, which is an equilibrium limit unable to capture many nonequilibrium properties such as energy dissipation and the flocking transition.
Given that the two-site AIM can be considered as a coarse-grained version of the full AIM,
it will be interesting to investigate what is the appropriate coarse-graining procedure that  preserves the dissipation characteristics, in particular, the cusped maximum behavior, and whether there is a scaling law for the dissipation as suggested by recent studies of general reaction networks~\cite{yu_inverse_2021,yu_state-space_2022,cocconi_scaling_2022}.
Finally, the energy-speed-sensitivity trade-off uncovered here may provide a useful perspective for understanding dynamics of natural flocks and designing optimal control of artificial flocks.

\begin{acknowledgments}
    This work is supported in part by National Institutes of Health Grant No.~R35GM131734 (to Y.~T.). Q.~Y.~acknowledges the IBM Exploratory Science Councils for a summer internship during which the work was finished.
    Q.~Y.~also acknowledges helpful discussions with D.~Zhang, Q.~Ouyang, C.~Tang, L.~Di Carlo, and C.~Wu.
\end{acknowledgments}

\appendix

\renewcommand{\theequation}{A\arabic{equation}}
\renewcommand{\thefigure}{A\arabic{figure}}
\setcounter{figure}{0}
\setcounter{equation}{0}

\section{Appendix A: Energy dissipation in the 2D AIM}
The dynamics of the 2D AIM is simulated using the random-sequential-update algorithm outlined in the original model~\cite{solon_flocking_2015}. The steady-state energy dissipation rate is obtained by computing the average energy dissipation rate of a sufficiently long trajectory~\cite{peliti_stochastic_2021},
\begin{equation}
    \dot{W} = \lim\limits_{\tau\to+\infty} \frac{1}{\tau}{\ln\frac{\mathcal{P}}{\mathcal{P}^R}} = \lim\limits_{\tau\to+\infty} \frac{1}{\tau} \sum_{i=0}^{M-1}\ln \frac{\tilde k_{i}}{\tilde k_{i}^R},
\end{equation}
where $\mathcal{P}$ and $\mathcal{P}^R$ are the probabilities of observing the forward and backward trajectories~\footnote{The spin variables do not change sign under time reversal.}. The trajectory contains $M$ transitions (flipping or hopping), with $\tilde k_{i}$ and $\tilde k_{i}^R$ being the forward and backward transition rates of the $i$th transition (for a single spin). Thus, the rate ratios for flipping and hopping reactions can be summed separately, which leads to the partition between alignment and motion decomposition.
For hopping, the log rate ratio $\ln(\tilde k_{i} / \tilde k_{i}^R)$ is simply $s\Delta x\ln\frac{1+\epsilon}{1-\epsilon}$, where $\Delta x=\pm1$ is the displacement in the $x$ direction. Hopping is along the bias when $s$ and $\Delta x$ have the same sign, which leads to positive dissipation. Conversely, hopping against the bias leads to negative dissipation. Hopping in the $y$ direction does not contribute to dissipation since forward and backward rates are equal. Therefore, the motion dissipation rate is:
\begin{align}
    \dot{W}_m = \lim\limits_{\tau\to+\infty} \frac{1}{\tau} \sum_{0<t<\tau} \Delta x s\ln\frac{1+\epsilon}{1-\epsilon} = Nv_0\ln\frac{1+\epsilon}{1-\epsilon},
    \label{Eq:appendix Wm}
\end{align}
where $v_0=2D\epsilon$ is the average speed along the bias.
Similarly, the forward and backward flipping rates are $\tilde k_{i}=\omega e^{-E_0sm/\rho}$ and $\tilde k_{i}^R=\omega e^{E_0s(m-2s)/\rho}$. The alignment dissipation is computed by summing the log ratios of the rates (Eq.~\ref{Eq: wdot flip definition}). A more detailed discussion can be found in the Supplemental Material.

\begin{figure}
    \includegraphics[width=\linewidth]{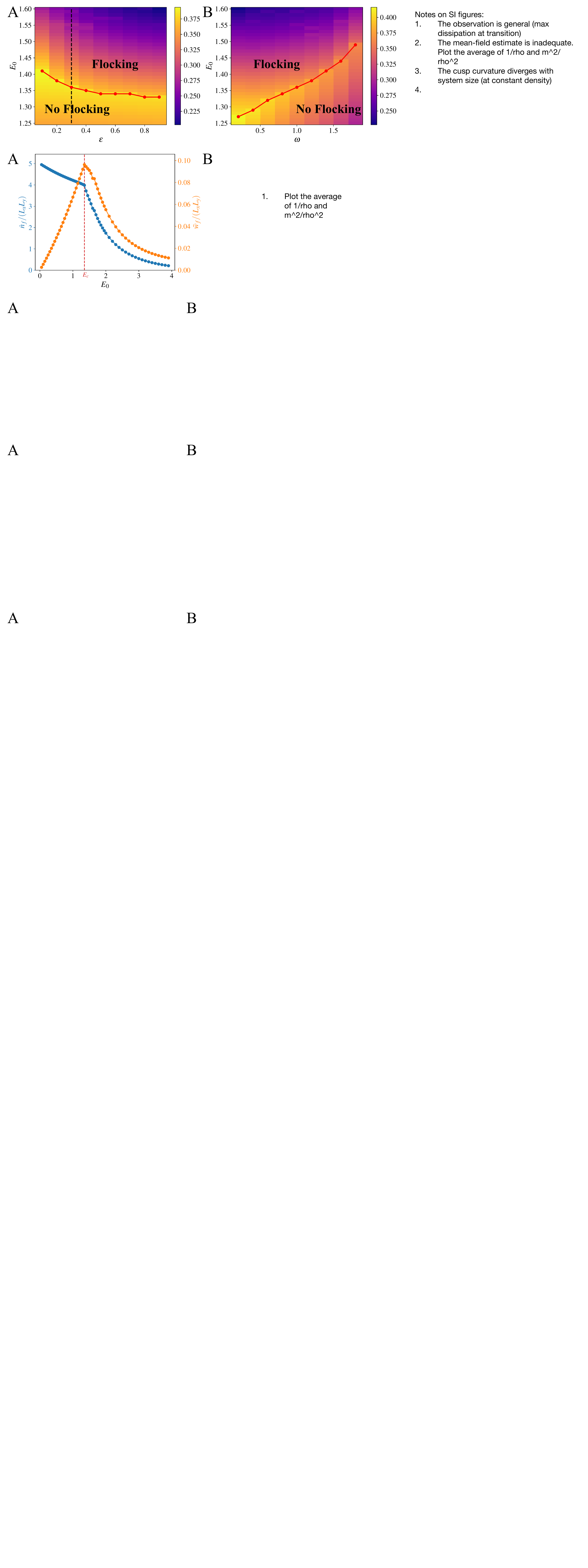}
    \caption{
        Generality of the alignment dissipation maximum at flocking transition.
        The heat maps show the alignment dissipation density $\dot{w}_a/(L_xL_y)$ for different combinations of (A) $(E_0,\epsilon)$ and (B) $(E_0, \omega)$.
        The red lines indicate the flocking transition as measured by the flocking velocity $v$. $D=1$, $L_x=300$, $L_y=100$, $\bar{\rho} = N/(L_xL_y)=5$. $\omega = 1$ for (A); $\epsilon=0.3$ for (B).
        The black dashed line in (A) shows the parameter values for Fig.~\ref{Fig1: full AIM dissipation}B.
    }
    \label{Fig: dissipation heatmaps}
\end{figure}

The generality of the alignment dissipation maximum at flocking transition shown in Fig.~\ref{Fig1: full AIM dissipation}B is confirmed by simulation using different values of $\epsilon$ (Fig.~\ref{Fig: dissipation heatmaps}A) and $\omega$ (Fig.~\ref{Fig: dissipation heatmaps}B). The red lines indicate the flocking transition $E_c$ measured by the flocking velocity $v$, and the alignment dissipation density $\dot{w}_a/(L_xL_y)$ is quantified by the heat maps.  In all cases studied, the dissipation maximum coincides with the flocking transition $E_c$, demonstrating generality of the result.

\renewcommand{\theequation}{B\arabic{equation}}
\renewcommand{\thefigure}{B\arabic{figure}}
\setcounter{figure}{0}
\setcounter{equation}{0}
\section{Appendix B: Analytical results in the two-site  and three-site AIM}
We start by computing $\dot{w}_a$ for the two-site AIM.
To obtain the steady-state distribution $P^s$, we decompose the linear operator $\mathcal{L}$ into $\mathcal{L}=D \mathcal{L}_1 + \omega \mathcal{L}_2$, where $\mathcal{L}_1$ captures hopping transitions and $\mathcal{L}_2$ captures flipping~\footnote{Their full expressions can be found in the SM.}. The steady-state condition becomes $(\mathcal{L}_1+\frac{\omega}{D}\mathcal{L}_2)P=0$, where the second term is treated as a perturbation for small $\omega/D$. To the leading order in $\omega/D$, $P$ can be written as
\begin{align}
    P = Q_0(a_0)\qty(p_0+\frac{\omega}{D} p_1) + \mathcal{O}\qty(\frac{\omega}{D})^2,
    \label{Eq:appendix twosite p}
\end{align}
where $p_0=2^{-N}\binom{a_0}{a_1}\binom{N-a_0}{b_1}$ is the solution to the hopping operator (i.e. $\mathcal{L}_1 p_0=0$), and $Q_0(a_0)$ captures the distribution of the total magnetization due to flipping. $Q_0$ and $p_1$ are determined by expanding $\mathcal{L}P=0$ to $\mathcal{O}(\frac{\omega}{D})$:
\begin{equation}
    \mathcal{L}_2(Q_0p_0) + \mathcal{L}_1(Q_0p_1) =0.
    \label{Eq:appendix twosite eq}
\end{equation}
First, we eliminate $\mathcal{L}_1$ by summing over $a_1$ and $b_1$, which leads to the steady-state condition for $Q_0$:
\begin{align}
    \sum_{a_1,b_1} \mathcal{L}_2(p_0Q_0)=0,\quad \forall a_0.
\end{align}
In the infinite $N$ limit, the solution is $Q_0(z) = e^{-F}$, where $z=a_0/N$ is the fraction of spin-up and $F$ is the effective free energy given by Eq.~\ref{Eq:free energy}. $p_1$ is determined by substituting the $Q_0$ solution into Eq.~\ref{Eq:appendix twosite eq}:
\begin{align}
    p_1=p_0\cdot\frac{E_0}{4}\psi_1\qty(z(1-z)N\qty(x-y)^2-1),
\end{align}
where $x=2a_1/a_0-1$ and $y=2b_1/(N-a_0)-1$, and
\begin{align}
    \psi_1 = z e^{E_0(1-2z)} +(1-z)e^{-E_0(1-2z)}.
\end{align}
The total steady-state energy dissipation (entropy production) rate is
\begin{equation}
    \dot{W}_\mathrm{tot} = \sum_{i<j}\qty(J_{i\to j} - J_{j \to i}) \ln \frac{k_{i\to j}}{k_{j \to i}},
\end{equation}
where the summation goes over all pairs of transitions $(i,j)$, which enables the decomposition into $\dot{W}_m$ and $\dot{W}_a$ by summing flipping and hopping transitions separately. For the motion dissipation, the calculation recovers Eq.~\ref{Eq:appendix Wm}. The alignment dissipation is given by the expectation value of the dissipation rate density $\sigma$ (defined in Eq.~\ref{Eq:wa twosite definition}):
\begin{widetext}
    \begin{align}
        \sigma= \frac{1}{2\omega E_0} \sum_{\mathrm{flip}} (J_+-J_-) \ln \frac{k_+}{k_-} = 2E_0 z(1-z)^2e^{E_0(2z-1)}\qty(4z(1-z)-\frac{\omega}{D}\frac{\psi_1}{2}) N(x-y)^2 + \mathcal{O}(x^4,y^4),
    \end{align}
\end{widetext}
where the higher-order terms in $x$ and $y$ are omitted since their expectation value vanishes in the infinite $N$ limit. Importantly, $\sigma$ vanishes exactly at the saddle point $(x,y,z)=(0,0,z^\star)$. Therefore, $\dot{w}_a$ comes from the second-order term $N(x-y)^2$.
This expansion directly relates $\dot{w}_a$ to the particle number fluctuations since $x$ and $y$ are the continuum versions of $a_1$ and $b_1$.
The saddle-point integral in the $(x,y)$ directions is done by averaging  $N(x-y)^2$ over $\qty(p_0+\frac{\omega}{D}p_1)$:
\begin{align}
    \expval{N(x-y)^2}
    =\frac{1}{z(1-z)}\qty(1+\frac{\omega E_0}{2D}\psi_1).
\end{align}
The integral in the $z$ direction is trivial since $z$ can simply take its saddle-point value. The result is
\begin{widetext}
    \begin{equation}
        {\dot{w}_a = \begin{cases}
                E_0\qty(1+\frac{\omega}{2D}(E_0-1)+\mathcal{O}\qty(\frac{\omega}{D})^2),\quad                                                                                                   & E_0<1. \\
                8E_0(z^\star)^{3/2}(1-z^\star)^{3/2}\qty(1+\frac{\omega}{2D}\qty(2E_0\sqrt{z^\star(1-z^\star)}-\frac{1}{2\sqrt{z^\star(1-z^\star)}})+\mathcal{O}\qty(\frac{\omega}{D})^2),\quad & E_0>1.
            \end{cases}}
        \label{Eq: wa two-site perturbation result}
    \end{equation}
\end{widetext}
These expressions explicitly demonstrate how the alignment dissipation depend on both $E_0$ and the relative timescale $\omega/D$. They are in good agreement with results obtained from the numerical solution of the master equation (see Fig.~S6 in the Supplemental Material). $\dot{w}_a$ exhibits a cusped maximum at $E_c=1$ regardless of $\omega/D$ because both the leading-order term and the correction term have discontinuous first derivatives there.
The $\mathcal{O}(\frac{\omega}{D})$ term is not included in the comparison with the 2D AIM (Fig.~\ref{Fig2: two-site}B) since the time needed to diffuse through the whole system is much longer than the timescale for flipping in the full model.

The three-site AIM can be solved by using the same method. The main difference is that the hopping operator $\mathcal{L}_1$ explicitly depends on the bias $\epsilon$, which enables us to capture the $\epsilon$ dependence of $\dot{w}_a$. Two new variables $a_2$ and $b_2$ are introduced for the number of $+$ and $-$ spins on site 2.
To the first order in $\omega/D$, the steady-state distribution is $P^s = Q_0(a_0)\qty(p_0+\frac{\omega}{D} p_1)$, where $p_0=3^{-N}\binom{a_0}{a_1}\binom{a_0-a_1}{a_2}\binom{b_0}{b_1}\binom{b_0-b_1}{b_2}$ is the solution to the hopping operator (i.e. $\mathcal{L}_1 p_0=0$). $Q_0=e^{-F}$ is the same as that of the two-site model, and so is the saddle point $z^\star$. The correction $p_1$ depends on both $E_0$ and $\epsilon$ with its full expression given in the Supplemental Material.
The alignment dissipation is calculated using the saddle point method, which involves expanding to the second order in particle numbers. The result is
\begin{widetext}
    \begin{equation}
        {\dot{w}_a = \begin{cases}
                2E_0\qty(1+\frac{\omega}{D}\frac{6+\epsilon^2}{3(3+\epsilon^2)}(E_0-1)+\mathcal{O}\qty(\frac{\omega}{D})^2),\quad                                                                                                                           & E_0<1, \\
                16E_0(z^\star)^{3/2}(1-z^\star)^{3/2}\qty(1+\frac{\omega}{D} \frac{6+\epsilon^2(2-4z^\star(1-z^\star))}{3(3+\epsilon^2)}\qty(2E_0\sqrt{z^\star(1-z^\star)}-\frac{1}{2\sqrt{z^\star(1-z^\star)}})+\mathcal{O}\qty(\frac{\omega}{D})^2),\quad & E_0>1.
            \end{cases}}
        \label{Eq: wa three-site perturbation result}
    \end{equation}
\end{widetext}
In addition to capturing the cusped maximum at the transition, the three-site result also reveals how $\dot{w}_a$ depends on $\epsilon$. It is in good agreement with numerical results obtained using the Gillespie algorithm~\cite{gillespie_exact_1977} (see Fig.~S7 in the Supplemental Material).

\bibliography{flocking}

\end{document}


\title{{
            Supplemental Material: Energy cost for flocking of active spins: the cusped dissipation maximum at the flocking transition
        }}

\author{Qiwei Yu}
\affiliation{IBM T.~J.~Watson Research Center, Yorktown Heights, NY 10598}
\affiliation{Lewis-Sigler Institute for Integrative Genomics, Princeton University, Princeton, NJ 08544}

\author{Yuhai Tu}
\affiliation{IBM T.~J.~Watson Research Center, Yorktown Heights, NY 10598}
\date{\today}

\maketitle
\tableofcontents

\section{Energy dissipation of the 2D active Ising model}
\subsection{Methods of numerical simulation and dissipation calculation}
Our simulation of the AIM completely follows refs.~\cite{solon_revisiting_2013,solon_flocking_2015}.
We use a random-sequential-update algorithm with time step $\Delta t = [4D+ e^{\beta E_0}]^{-1}$.
For each time interval $\Delta t$, the system is updated $N$ times, where $N$ is the number of particles.
For each update, a particle is chosen at random and updated with one of the outcomes: flipping with probability $\omega e^{-\beta E_0 s m_{i,j}/\rho_{i,j}} \Delta t$, moving to right/left with probability $D(1\pm s\epsilon)\Delta t$, moving up/down with probability $D\Delta t$, or doing nothing with probability $1-\qty(4D+\omega e^{-\beta E_0 s m_{i,j}/\rho_{i,j}})\Delta t$. The inverse temperature is set to $\beta=1$ throughout.

For Fig.~1 of the main text, each simulation is run for 3.6 million time steps. During each time step, the system is updated $N$ times. The mean velocity and dissipation are computed for each time step (which includes $N$ updates). Fig.~1 shows these quantities averaged over all time steps, and the error bars are the standard deviation over all time steps.

The state variable of the system is the occupancy of all the sites: $\mathbf{n} = \qty(n^+_{1,1},n^-_{1,1},n^+_{1,2},n^-_{1,2},\dots, n^+_{L_x,L_y}, n^-_{L_x,L_y})$. These states span a high-dimensional state space with transition rate from $\mbn$ to $\mbn'$ denoted by $k_{\mbn\to\mbn'}$. Our objective is to calculate the dissipation from the dynamics represented by a trajectory (time series of states) :
\begin{equation}
    \mbn(t_0),\mbn(t_1),\mbn(t_2),\dots,\mbn(t_m),\quad
    t_0<t_1<t_2<\cdots<t_m.
\end{equation}
where $\mbn(t_0)$ is the initial state at $t_0$;
${t_1,t_2,t_3,\dots,t_{m}}$ are the time of transition events. For the sake of simplicity, $\mbn(t_i)$ will be denoted by $\mbn_i$. We also define $k_{\mbn_i}^\mathrm{out}$ as the total rate of exiting state $\mbn_i$.

The (average) energy dissipation rate of the trajectory is:
\begin{equation}
    \dot{W} = \lim\limits_{t\to+\infty} \frac{1}{t}\expval{\ln\frac{\mathcal{P}}{\mathcal{P}^R}},
\end{equation}
where $\mathcal{P}$ is the probability of observing the forward trajectory, and $\mathcal{P}^R$ is the probability of observing the reverse trajectory.
$\expval{\cdot}$ is the averaging over noise realizations, which will be omitted by assuming ergodicity. The spin variables do not change sign under time reversal.

The probability of the forward trajectory is:
\begin{align}
    \mathcal{P} & =P(\mbn_0) e^{-(t_1-t_0)k_{\mbn_0}^\mathrm{out}} k_{\mbn_0\to\mbn_1} e^{-(t_2-t_1)k_{\mbn_1}^\mathrm{out}}k_{\mbn_1\to\mbn_2} \cdots e^{-(t_m-t_{m-1})k_{\mbn_{m-1}}^\mathrm{out}}k_{\mbn_{m-1}\to\mbn_m} \\
                & =
    P(\mbn_0) e^{-\sum_{i=0}^{m-1}(t_{i+1}-t_i)k^\mathrm{out}_{\mbn_i}}\prod_{i=0}^{m-1}k_{\mbn_i\to\mbn_{i+1}}
\end{align}
The probability of the reverse trajectory is:
\begin{align}
    \mathcal{P}^R & = P(\mbn_m)
    k_{\mbn_{m}\to\mbn_{m-1}}
    e^{-(t_m-t_{m-1})k_{\mbn_{m-1}}^\mathrm{out}}
    \cdots k_{\mbn_1\to\mbn_0}e^{-(t_1-t_0)k_{\mbn_0}^\mathrm{out}}                                                                \\
                  & =P(\mbn_m) e^{-\sum_{i=0}^{m-1}(t_{i+1}-t_i)k^\mathrm{out}_{\mbn_i}}\prod_{i=0}^{m-1}k_{\mbn_{i+1}\to\mbn_{i}}
\end{align}
Therefore,
\begin{align}
    \ln \frac{\mathcal{P}}{\mathcal{P}^R} = \ln\frac{P(\mbn_0)}{P(\mbn_m)} + \sum_{i=0}^{m-1}\ln \frac{k_{\mbn_i\to\mbn_{i+1}}}{k_{\mbn_{i+1}\to\mbn_{i}}}.
\end{align}
The first term is bounded, so its contribution to $\dot{W}$ vanishes in the $t\to\infty$ limit. The second term is summed over all the transitions (flipping or hopping events) that occur in time $t$.
Therefore, the dissipation rate reads
\begin{equation}
    \dot{W} = \lim\limits_{t\to+\infty} \frac{1}{t} \sum_{i=0}^{m-1}\ln \frac{k_{\mbn_i\to\mbn_{i+1}}}{k_{\mbn_{i+1}\to\mbn_{i}}},
\end{equation}
where the summation goes over all the flipping and hopping events during time $t$.
The rate $k_{\mbn_i\to\mbn_{i+1}}$ can be further decomposed into the bare rate $\tilde k_{\mbn_i\to\mbn_{i+1}}$, which is the rate for a single particle, and the number of spins in the same configuration $n_{x,y}^\pm$. For example, for a flipping transition that flips $s$ to $(-s)$ on site $(i,j)$, we have
\begin{align}
    \tilde k_{\mbn\to\mbn'} = \omega e^{-E_0sm_{i,j}/\rho_{i,j}},\quad
    k_{\mbn\to\mbn'}= n_{i,j}^s \tilde k_{\mbn\to\mbn'} = \frac{\rho_{i,j}+sm_{i,j}}{2} \omega e^{-E_0sm_{i,j}/\rho_{i,j}}.
\end{align}
For the sake of simplicity, the bare rates $\tilde k_{\mbn_i\to\mbn_{i+1}}$ and $\tilde k_{\mbn_{i+1}\to\mbn_{i}}$ are denoted by $\tilde k_{i}$ and $\tilde k_{i}^R$ in Appendix A.

For a flipping event $\mbn\to\mbn'$ that flips a spin from $s$ to $(-s)$ on site $(i,j)$, the rate ratio reads
\begin{equation}
    \qty(\frac{k_{\mathbf{n}\to \mathbf{n'}}}{k_{\mathbf{n'} \to \mathbf{n}}})_{+s \to -s} = \frac{\rho_{i,j}+sm_{i,j}}{\rho_{i,j}-sm_{i,j}+2}\exp(- 2s\frac{m_{i,j}-s}{\rho_{i,j}}E_0) = \frac{\rho_{i,j}+sm_{i,j}}{\rho_{i,j}-sm_{i,j}+2}\exp(2E_0\frac{1-m_{i,j}s}{\rho_{i,j}}),
\end{equation}
where $m_{i,j}$ and $\rho_{i,j}$ are the local magnetization and density in state $\mbn$ (i.e., before flipping).

For hopping events in the $x$ direction, the rate ratio reads
\begin{equation}
    \qty(\frac{k_{\mathbf{n}\to \mathbf{n'}}}{k_{\mathbf{n'} \to \mathbf{n}}})_{(s,i,j) \to (s,i+\Delta x,j)} = \frac{\rho_{i,j} + sm_{i,j}}{\rho_{i+\Delta x,j} + sm_{i+\Delta x,j}+2} \cdot  \frac{1+\Delta xs\epsilon}{1-\Delta xs\epsilon},
\end{equation}
where spin $s$ moves $\Delta x=\pm1$ in the $x$ direction when going from state $\mbn$ to $\mbn'$.

For hopping events in the $y$ direction, the rate ratio reads
\begin{equation}
    \qty(\frac{k_{\mathbf{n}\to \mathbf{n'}}}{k_{\mathbf{n'} \to \mathbf{n}}})_{(s,i,j) \to (s,i,j+\Delta y)} = \frac{\rho_{i,j} + sm_{i,j}}{\rho_{i,j+\Delta  y} + sm_{i,j+\Delta y}+2},
\end{equation}
where spin $s$ moves $\Delta y=\pm1$ in the $y$ direction when going from state $\mbn$ to $\mbn'$.

The total dissipation is the sum of contributions from all three types of events listed above. Further, it can be decomposed into an energy term ${W}_1(t)$ and an entropy term ${W}_2(t)$:
\begin{align}
    \dot{W} = \lim\limits_{t\to+\infty} \frac{W_1(t)+W_2(t)}{t}
\end{align}
The energy term is
\begin{equation}
    W_1(t) = 2E_0\sum_{\mathrm{flip\ } s}\frac{1-ms}{\rho}
    + \sum_{\mathrm{hop\ } (s,\Delta x)} \Delta x s  \ln\qty(\frac{1+\epsilon}{1-\epsilon}).
\end{equation}
The entropy term is
\begin{equation}
    \begin{aligned}
        W_2(t) = \sum_{\mathrm{flip\ } s}\ln\frac{\rho_{i,j}+sm_{i,j}}{\rho_{i,j}-sm_{i,j}+2}
         & + \sum_{\mathrm{hop\ } (s,\Delta x)}  \ln\qty(\frac{\rho_{i,j} + sm_{i,j}}{\rho_{i+\Delta x,j}	 + sm_{i+\Delta x,j}+2}) + \sum_{\mathrm{hop\ } (s,\Delta y)}  \ln\qty(\frac{\rho_{i,j} + sm_{i,j}}{\rho_{i,j+\Delta y} + sm_{i,j+\Delta y}+2}).
    \end{aligned}
\end{equation}
The entropy term $W_2(t)$ is associated with the change in the total entropy $S$. Namely, $W_2(t) = - \Delta S = S(0)-S(t)$, where
\begin{equation}
    S = \ln \frac{N!}{n^+_{1,1}!n^-_{1,1}!n^+_{1,2}!n^-_{1,2}!\cdots n^+_{L_x,L_y}! n^-_{L_x,L_y}!}.
\end{equation}
$S$ is always bounded, so the dissipation contribution from the entropy term vanishes in the long time limit:
\begin{equation}
    \dot{W}_2 = \lim\limits_{t\to+\infty} \frac{W_2(t)}{t} = \lim\limits_{t\to+\infty} \frac{S(0)-S(t)}{t} =0.
\end{equation}
The energy term, which is equal to the total dissipation rate, includes contribution from hopping events, which we call the motion dissipation $\dot{W}_m$, and contribution from flipping events, which we call the alignment dissipation $\dot{W}_a$:
\begin{equation}
    \dot{W} = \dot{W}_1= \lim\limits_{t\to+\infty} \frac{W_1(t)}{t} = \dot{W}_{a} + \dot{W}_{m} = 2E_0 \lim\limits_{t\to+\infty} \frac{1}{t}\sum_{\mathrm{flip\ } s}\frac{1-ms}{\rho} + \ln\qty(\frac{1+\epsilon}{1-\epsilon})
    \lim\limits_{t\to+\infty} \frac{1}{t}\sum_{\mathrm{hop\ } (s,\Delta x)} \Delta x s  .
\end{equation}
The motion dissipation rate is
\begin{equation}
    \dot{W}_{m} =  \lim\limits_{t\to+\infty} \frac{1}{t} \sum_{\mathrm{hop\ } (s,\Delta x)} \Delta x s  \ln\qty(\frac{1+\epsilon}{1-\epsilon}) = 2D\epsilon N  \ln\qty(\frac{1+\epsilon}{1-\epsilon}),
\end{equation}
which is required to generate self-propulsion.
The alignment contribution is
\begin{equation}
    \dot{W}_{a} =  2E_0 \lim\limits_{t\to+\infty} \frac{1}{t}\sum_{\mathrm{flip\ } s}\frac{1-m_{i,j}s}{\rho_{i,j}},
\end{equation}
which is the extra energy cost needed to generate flocking order among self-propelled particles.

\subsection{Decomposing the alignment dissipation rate into frequency and cost}
\begin{figure}
    \includegraphics[width=0.8\linewidth]{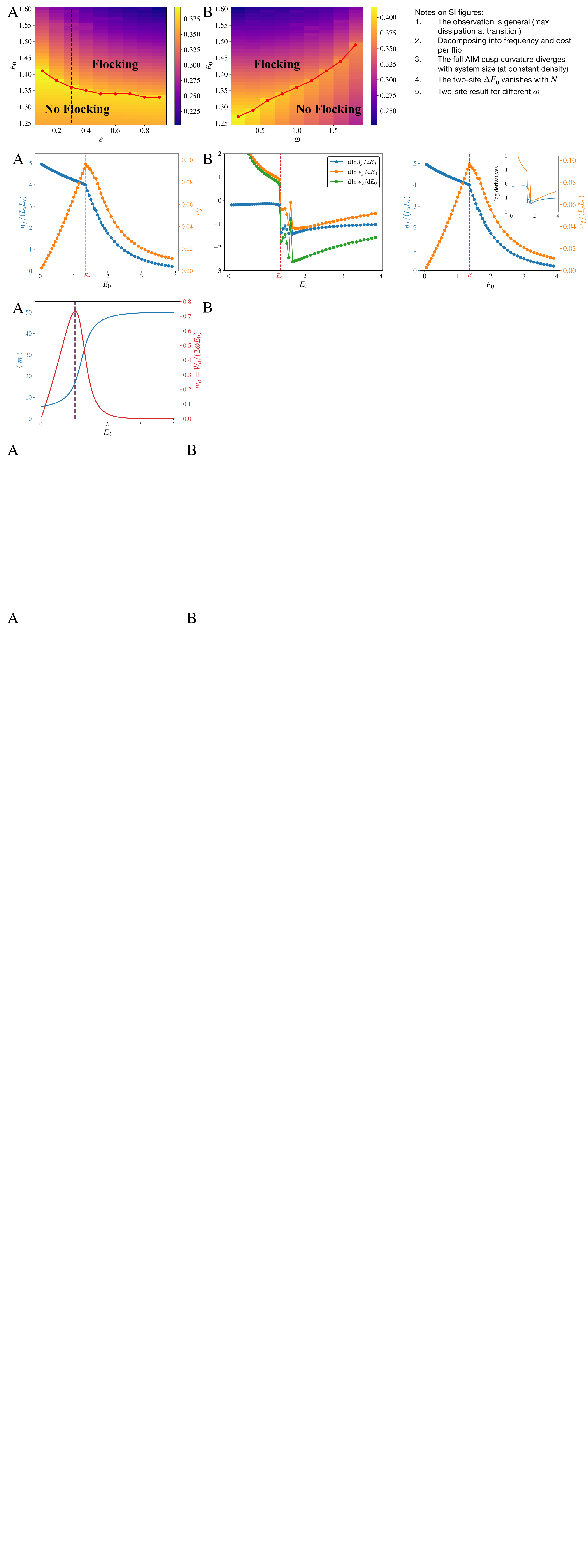}
    \caption{
        (A) The flipping frequency $\dot{n}_f$ (normalized by the lattice size $L_xL_y$) and the energy dissipation per flipping event $\bar{w}_f$ for different $E_0$. $E_c$ is the flocking transition point.
        (B) The log derivatives $\dv{\ln\dot{n}_f}{E_0}$, $\dv{\ln\bar w_f}{E_0}$, and $\dv{\ln\dot{w}_a}{E_0}$ all exhibit discontinuity at $E_c$.
        $D=\omega=1$, $L_x=300$, $L_y=100$, $\epsilon=0.3$.
    }
    \label{Fig: nf wf}
\end{figure}

The alignment dissipation can be decomposed by $\dot{w}_a = \dot{n}_f \bar{w}_f$, where $\dot{n}_f$ is the frequency of flipping events, and $\bar{w}_f$ is the average energy dissipation due to one flipping event. As shown in Fig.~\ref{Fig: nf wf}A, $\dot{n}_f$ decreases monotonically with $E_0$ while $\bar{w}_f$ peaks at the flocking transition $E_c$, but they both have discontinuous first derivatives at $E_c$. To decompose the discontinuity at $E_c$, we consider the log derivatives:
\begin{equation}
    \dv{\ln\dot{w}_a}{E_0} = \dv{\ln\dot{n}_f}{E_0} + \dv{\ln\bar w_f}{E_0}.
\end{equation}
Fig.~\ref{Fig: nf wf}B shows that both terms on the right hand side experience a sudden decrease at $E_c$, thereby contributing to the cusp behavior of $\dot{w}_a$.

\subsection{The cusp emerges in sufficiently large systems}
\begin{figure}
    \includegraphics[width=0.4\linewidth]{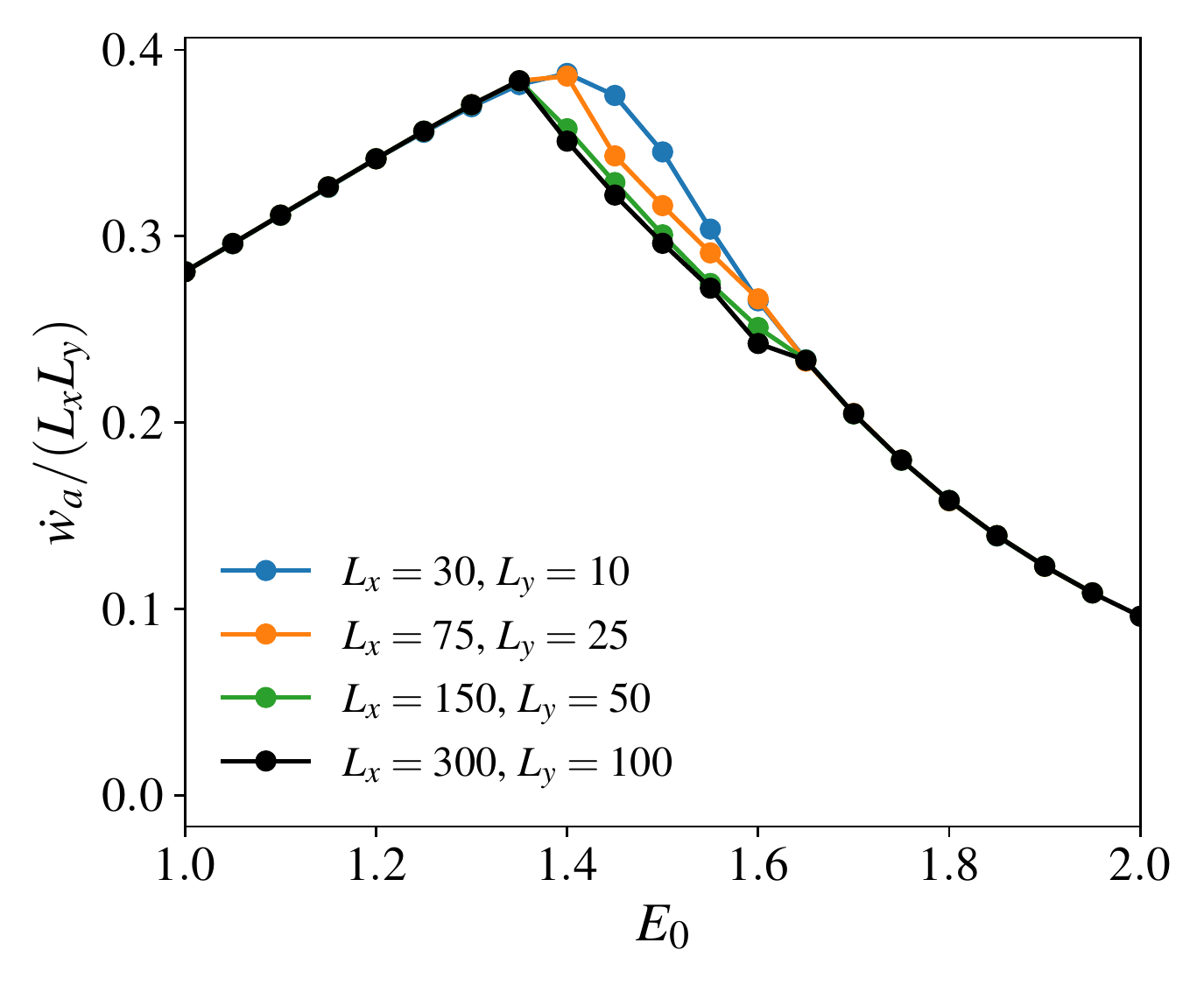}
    \caption{
        The alignment dissipation near the flocking transition for four different system sizes $(L_x,L_y)$ (see legend). The black curve is the system size used in this study.  $D=\omega=1$, $\bar{\rho}=5$, $\epsilon=0.3$.
    }
    \label{Fig: LxLy cusp}
\end{figure}

In the two-site model, the cusp emerges in the thermodynamic limit $N\to\infty$, which is demonstrated by showing the divergence of the peak curvature $\pdv[2]{\dot{w}_a}{E_0}$ with $N$ (Fig.~2B inset of the main text).
In the full AIM, the cusp also only appears when the system is sufficiently large.
We demonstrate this by simulating the AIM with different sizes (as shown in Fig.~\ref{Fig: LxLy cusp}).
For a small system (blue), the dissipation maximum is smooth with continuous derivatives (i.e. analytic).
As the system size increases (orange and green), the cusp emerges and the first derivative becomes discontinuous. {The convergence of curves indicates that the system size used in the main text ($L_x=300$, $L_y=100$) is sufficiently large.}

\subsection{The case of unbiased diffusive motion $\epsilon=0$}

\begin{figure}[h]
    \centering
    \includegraphics[width=\textwidth]{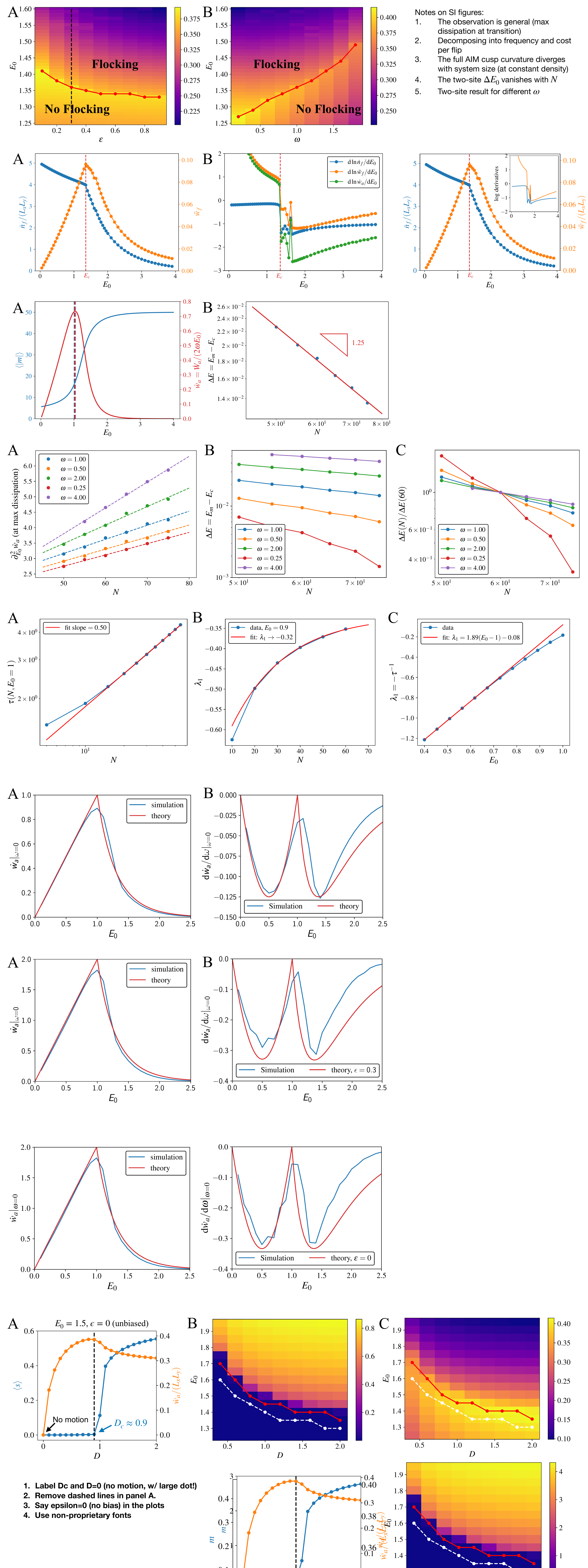}
    \caption{Average spin ${\expval{s}}$ and alignment dissipation density $\dot{w}_a/(L_xL_y)$ for $\epsilon=0$.
        (A) shows average spin and alignment dissipation for $E_0=1.5$. The blacked dashed line indicates $D_c$, which is the minimum motion needed to generate long-range order.
        (B) and (C) show average spin and dissipation for different $E_0$ and $D$, where the red line is the transition (critical) line determined by thresholding $\expval{s}$. For comparison, the white dashed line indicates the transition line for $\epsilon=0.1$, showing that the effect of the bias ($\epsilon$) is to shift the critical line $D_c(E_0)$ without changing the general behaviors.}
    \label{fig:unbiased}
\end{figure}

In the special case with $\epsilon=0$, long-range order can still exist when the diffusion constant $D$ is larger than a critical value $D_c$, even though the flocking speed vanishes $v=0$. However, the dissipation rate has the same cusped maximum behavior at the transition point as for the case with biased motion (see Fig.~\ref{fig:unbiased}). Note that the AIM system is out of equilibrium as long as $D\ne 0$.

\section{The two-site active Ising model}

\subsection{Master equation}
The two-site model consists of $N$ active spins on a lattice of size $L_x=2$ and $L_y=1$ with periodic boundary conditions. The particle dynamics follow those of the full AIM.

The state of the system is completely described by three variables (degrees of freedom) $(a_0,a_1,b_1)$. $a_0$ is the total number of spins up; $a_1$ and $b_1$ are the number of spins up/down on site 1. The total magnetization is $m=2a_0-N$. The number of spins up/down on site 2 is given by $a_2=a_0-a_1$ and $b_2=N-a_0-b_1$.
With these variables, we can write down the reaction rates for all possible hopping and flipping events. This leads to the Master Equation:
\begin{equation}
    \begin{aligned}
         & \dv{P(a_0,a_1,b_1)}{t} =  (a_1+1)\cdot 2D \cdot P(a_0,a_1+1,b_1) + (b_1+1)\cdot 2D \cdot P(a_0,a_1,b_1+1)                                                                                                                            \\
         & + (a_0-a_1+1)\cdot 2D \cdot P(a_0,a_1-1,b_1) + (N-a_0-b_1+1)\cdot 2D \cdot P(a_0,a_1,b_1-1)                                                                                                                                          \\
         & +(a_1+1)\cdot \omega e^{-E_0\frac{a_1-b_1+2}{a_1+b_1}} \cdot P(a_0+1,a_1+1,b_1-1) + (b_1+1)\cdot \omega e^{E_0\frac{a_1-b_1-2}{a_1+b_1}}  \cdot P(a_0-1,a_1-1,b_1+1)                                                                 \\
         & +(a_0-a_1+1)\cdot \omega e^{-E_0\frac{2a_0-N-a_1+b_1+2}{N-a_1-b_1}} \cdot P(a_0+1,a_1,b_1) + (N-a_0-b_1+1)\cdot \omega e^{E_0\frac{2a_0-N-a_1+b_1-2}{N-a_1-b_1}}  \cdot P(a_0-1,a_1,b_1)                                             \\
         & -\qty[2ND+a_1\omega e^{-E_0\frac{a_1-b_1}{a_1+b_1}}+b_1 \omega e^{E_0\frac{a_1-b_1}{a_1+b_1}}+(a_0-a_1)\omega e^{-E_0\frac{2a_0-N-a_1+b_1}{N-a_1-b_1}}  + (N-a_0-b_1) \omega e^{E_0\frac{2a_0-N-a_1+b_1}{N-a_1-b_1}}]P(a_0,a_1,b_1).
    \end{aligned}
    \label{Eq:two-site master equation}
\end{equation}
On the right hand side, the first four terms describe hopping; the next four terms describe flipping; the rest of the terms are the reverse reactions. We can write the master equation in a more compact form:
\begin{equation}
    \dv{P}{t} = \mathcal{L}\cdot P,
\end{equation}
where the matrix $\mathcal{L}$ consists of the transition rates.

Physical observables can be determined by averaging over the steady-state distribution:
\begin{equation}
    \expval{A} = \sum_{a_0,a_1,b_1}A(a_0,a_1,b_1)P(a_0,a_1,b_1) = \sum  A\cdot P.
\end{equation}
The probability is normalized by $\expval{1}=1$.
For example, the average magnetization $\expval{{m}}$ is obtained by
\begin{equation}
    \expval{{m}} = \sum_{a_0,a_1,b_1}\qty(2a_0-N)P(a_0,a_1,b_1) = \sum_{a_0}\qty(2a_0-N)\sum_{a_1,b_1}P(a_0,a_1,b_1).
\end{equation}
The free energy landscape for magnetization (plotted in Fig.~2A of the main text) is given by
\begin{equation}
    F(m) = -\ln P(m) = -\ln \sum_{a_1,b_1}P\qty(\frac{N+m}{2},a_1,b_1).
\end{equation}
The total energy dissipation reads:
\begin{equation}
    \dot{W}_\mathrm{tot} = \sum_{i<j}\qty(J_{i\to j} - J_{j \to i}) \ln \frac{J_{i\to j}}{J_{j \to i}}= \sum_{i<j}\qty(J_{i\to j} - J_{j \to i}) \ln \frac{k_{i\to j}}{k_{j \to i}}.
\end{equation}
where $\sum_{i<j}$ sums over all pairs of (forward and backward) reactions connecting microscopic states, and $J_{i\to j} = k_{i\to j} P_i$ is the steady-state probability flux from state $i$ to state $j$. The notation $i<j$ avoids double counting.
Notably, the reactions of hopping to the left and to the right must be treated separately for thermodynamic purposes. Although they arrive at the same site due to the periodic boundary condition, their energy dissipation has opposite signs (i.e. $\ln\frac{1+\epsilon}{1-\epsilon}$ along the bias and $\ln\frac{1-\epsilon}{1+\epsilon}$ against the bias).

The alignment dissipation is calculated by summing up the dissipation on all reaction links associated with flipping:
\begin{equation}
    \dot{w}_{a} = \frac{\dot{W}_{a}}{2\omega E_0}  = \frac{1}{2\omega E_0}\sum_{\expval{i,j}\in \mathcal{E}_\mathrm{flip}}\qty(J_{i\to j} - J_{j \to i}) 2E_0\frac{1-m_is}{\rho_i} = \frac{1}{\omega} \sum_{\expval{i,j}\in \mathcal{E}_\mathrm{flip}}\qty(J_{i\to j} - J_{j \to i}) \frac{1-m_is}{\rho_i},
\end{equation}
where $\mathcal{E}_\mathrm{flip}$ is the set of all links describing flipping. The transition from state $i$ to state $j$ flips $s$ to $(-s)$. $m_i$ and $\rho_i$ are the local magnetization and density in state $i$ (i.e. prior to flipping).  The total dissipation is the sum of alignment and motion dissipation:
\begin{equation}
    \dot{W}_\mathrm{tot} = \sum_{i<j}\qty(J_{i\to j} - J_{j \to i}) \ln \frac{k_{i\to j}}{k_{j \to i}}=
    \dot{W}_a + \dot{W}_m= 2\omega E_0 \dot{w}_a+ 2D\epsilon N\ln\frac{1+\epsilon}{1-\epsilon}.
\end{equation}

\subsection{The maximum of the alignment dissipation coincides with the flocking transition}
In this section, we compare the flocking transition point $E_c$, which is determined from the curvature of the free energy landscape $F''(0)=0$, and the maximum of the alignment dissipation $E_m=\argmax_{E_0}\dot{w}_a$.

\begin{figure}
    \includegraphics[width=0.8\linewidth]{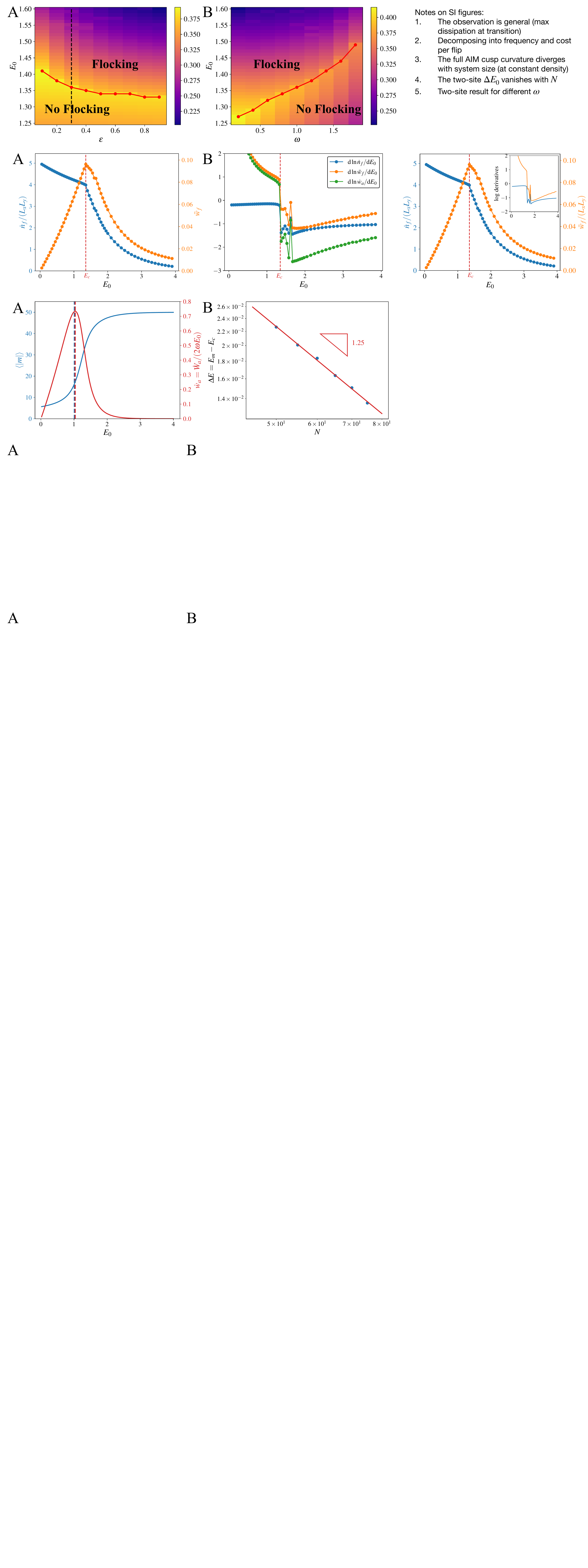}
    \caption{
        (A) The order parameter $\expval{\abs{m}}$ and alignment dissipation $\dot{w}_a$ for $N=50$.
        The blue dashed line is the flocking transition point $E_c$ determined from the free energy landscape. The red dashed line is the alignment dissipation maximum $E_m$.
        (B) The difference $\Delta E = E_m-E_c$ decreases with $N$.
    }
    \label{Fig: Delta E0}
\end{figure}

Fig.~\ref{Fig: Delta E0}A plots the typical behavior of $\expval{\abs{m}}$ and $\dot{w}_a$ as function of $E_0$ for a finite $N$. Here, $\expval{\abs{m}}$ is used as the order parameter since the absence of spontaneous symmetry breaking leads to $\expval{m}=0$ for finite $N$.
The flocking transition point $E_c$ (blue dashed line) is indeed associated with the rapid increase of $\expval{\abs{m}}$, and it is close to $E_m$.

The difference between $E_c$ and $E_m$ is further quantified in Fig.~\ref{Fig: Delta E0}B, which shows that $\Delta E = E_m-E_c$ decreases with $N$ following a power law. In the infinite $N$ limit, $\Delta E$ vanishes and the dissipation maximum exactly coincides with the flocking transition point.

\subsection{Solution in the infinite $N$ limit.}
In this section, we obtain the steady-state probability distribution of the two site AIM in the limit of infinite $N$.
We start with the Master Equation of the two-site model (Eq.~\ref{Eq:two-site master equation}).
We define $\mathcal{L}_1$ and $\mathcal{L}_2$ as the operators for hopping and flipping, respectively:
\begin{align}
    \frac{1}{2}\mathcal{L}_1P = & (a_1+1)\cdot  P(a_0,a_1+1,b_1) + (b_1+1)\cdot  P(a_0,a_1,b_1+1)                                                                                                                                        \\
                                & + (a_0-a_1+1)\cdot   P(a_0,a_1-1,b_1) + (N-a_0-b_1+1)\cdot  P(a_0,a_1,b_1-1)-NP(a_0,a_1,b_1).                                                                                                          \\
    \mathcal{L}_2P =            & (a_1+1)\cdot  e^{-E_0\frac{a_1-b_1+2}{a_1+b_1}} \cdot P(a_0+1,a_1+1,b_1-1) + (b_1+1)\cdot e^{E_0\frac{a_1-b_1-2}{a_1+b_1}}  \cdot P(a_0-1,a_1-1,b_1+1)                                                 \\
                                & +(a_0-a_1+1)\cdot  e^{-E_0\frac{2a_0-N-a_1+b_1+2}{N-a_1-b_1}} \cdot P(a_0+1,a_1,b_1) + (N-a_0-b_1+1)\cdot e^{E_0\frac{2a_0-N-a_1+b_1-2}{N-a_1-b_1}}  \cdot P(a_0-1,a_1,b_1)                            \\
                                & -\qty[a_1 e^{-E_0\frac{a_1-b_1}{a_1+b_1}}+b_1 e^{E_0\frac{a_1-b_1}{a_1+b_1}}+(a_0-a_1) e^{-E_0\frac{2a_0-N-a_1+b_1}{N-a_1-b_1}}  + (N-a_0-b_1) e^{E_0\frac{2a_0-N-a_1+b_1}{N-a_1-b_1}}]P(a_0,a_1,b_1).
\end{align}
The steady-state solution of the Master Equation satisfies
\begin{equation}
    \qty(\mathcal{L}_1 + \frac{\omega}{D} \mathcal{L}_2) P =0,\quad \forall (a_0,a_1,b_1).
\end{equation}

In the fast diffusion limit ($\frac{\omega}{D} \ll 1$), the leading order solution is
\begin{equation}
    P_0 = Q_0(a_0) p_0(a_0,a_1,b_1)=Q_0(a_0)\binom{a_0}{a_1}\binom{N-a_0}{b_1},
\end{equation}
where $p_0=\binom{a_0}{a_1}\binom{N-a_0}{b_1}$ is the solution for the hopping operator (i.e. $\mathcal{L}_1 p_0=0$), and $Q_0$ captures the dependence on $a_0$ which results from flipping.
For any given $a_0$, fast diffusion means that the spins follow binomial distribution on the two sites, which is captured by $p_0$. The steady-state distribution of $a_0$ is captured by $Q_0$, which is determined by $\sum_{a_1,b_1}\mathcal{L}_2(Q_0p_0)=0$ ($\forall a_0$).

For finite $\frac{\omega}{D}$, the perturbed solution is
\begin{equation}
    P = \qty(Q_0(a_0) + Q_1(a_0))\qty(p_0+\frac{\omega}{D} p_1 + \qty(\frac{\omega}{D})^2 p_2 +\cdots)
\end{equation}
where $Q_1$ is higher order in $\omega$ ($\lim\limits_{\omega\to 0} Q_1(a_0)=0$).
After substituting this into the steady-state condition, we find the leading order equation to be
\begin{equation}
    Q_0\mathcal{L}_1 p_1 + \mathcal{L}_2\qty(p_0Q_0) =0
    \label{Eq: p1 operator equation}
\end{equation}
which gives $p_1$. Similar analysis can be applied to obtain equations for $Q_1$, $p_2$, and so on, but they prove to be unnecessary as only $p_1$ is needed for calculating the leading-order correction to the energy dissipation.

\subsubsection{Solution of $Q_0$}
To determine $Q_0$, we substitute $P=p_0Q$ into the flipping operator:
\begin{equation}
    \begin{aligned}
        \mathcal{L}_2(Qp_0) =
         & \frac{(a_0+1)p_0Q(a_0+1)}{b_0}\qty[b_1e^{-E_0\frac{a_1-b_1+2}{a_1+b_1}}+(b_0-b_1)\cdot e^{-E_0\frac{2a_0-N-a_1+b_1+2}{N-a_1-b_1}} ]                                                                 \\
         & +\frac{(b_0+1)p_0Q(a_0-1)}{a_0}\qty[a_1e^{E_0\frac{a_1-b_1-2}{a_1+b_1}}+(a_0-a_1)\cdot e^{E_0\frac{2a_0-N-a_1+b_1-2}{N-a_1-b_1}} ]                                                                  \\
         & - \qty(a_1e^{-E_0\frac{a_1-b_1}{a_1+b_1}}+b_1  e^{E_0\frac{a_1-b_1}{a_1+b_1}}+(a_0-a_1) e^{-E_0\frac{2a_0-N-a_1+b_1}{N-a_1-b_1}}  + (N-a_0-b_1)  e^{E_0\frac{2a_0-N-a_1+b_1}{N-a_1-b_1}})p_0Q(a_0).
    \end{aligned}
    \label{Eq: L2p0Q definition}
\end{equation}
To solve for $Q_0$ in the large $N$ limit, we take the equation to the continuum limit by substituting the $a$ and $b$ variables by $x,y,z$:
\begin{equation}
    a_0 = Nz,\quad a_1 = \frac{a_0}{2}(1+x),\quad b_1 = \frac{N-a_0}{2}(1+y).
\end{equation}
The distribution $p_0=\binom{a_0}{a_1}\binom{N-a_0}{b_1}$ gives moments $\expval{x}_0=\expval{y}_0=0$, $\expval{x^2}_0 = \frac{1}{Nz}$, and $\expval{y^2}_0 = \frac{1}{N(1-z)}$, where $\expval{}_0$ denotes averaging over $p_0$. Let $\lambda_i = m_i/\rho_i$ be the average spin on site $i$.
To the leading order in $N$, we have $\lambda_1 = \frac{a_1-b_1}{a_1+b_1} = 2z-1 +O(x,y)$, and $\lambda_2=\frac{2a_0-N-a_1+b_1}{N-a_1-b_1} = 2z-1 + O(x,y)$, where $O(x,y)\sim O(N^{-1/2})$ are higher order terms in $N$. The flipping operator simplifies to (to the leading order in $N$):
\begin{align}
    \frac{\mathcal{L}_2(Qp_0)}{p_0}=
      & (a_0+1)Q(a_0+1) e^{-E_0(2z-1)} + (N-a_0+1)Q(a_0-1) e^{E_0(2z-1)}  - [a_0e^{-E_0(2z-1)}+(N-a_0)e^{E_0(2z-1)}]Q(a_0) \\
    = & \pdv{}{a_0}\qty[(a_0+1)Q(a_0+1) e^{-E_0(2z-1)}-(N-a_0)e^{E_0(2z-1)}Q(a_0)] = -\pdv{J_\mathrm{flip}}{a_0},
\end{align}
where $J_\mathrm{flip}=(N-a_0)e^{E_0(2z-1)}Q(a_0)-(a_0+1)Q(a_0+1) e^{-E_0(2z-1)}$ is the total probability flux for flipping a spin up. This flux vanishes at the steady state, which leads to:
\begin{equation}
    \frac{Q(a_0+1)}{Q(a_0)} = \frac{N-a_0}{a_0} e^{2E_0(2z-1)} = \frac{1-z}{z}  e^{2E_0(2z-1)}.
\end{equation}
Following the main text, the effective free energy $F$ is defined as
\begin{equation}
    F = -\ln Q.
\end{equation}
Therefore,
\begin{align}
    \pdv{F}{z}              & = -N\ln\frac{Q(a_0+1)}{Q(a_0)} = -N\qty(\ln\frac{1-z}{z}+ 2E_0(2z-1)), \\
    \Rightarrow \frac{F}{N} & = z\ln z + (1-z) \ln(1-z) + 2E_0z(1-z),
\end{align}
which recovers the effective free energy landscape reported in the main text (Eq.~3). The distribution $Q_0=e^{-F}$ is localized near the minima (saddle points) of the effective free energy. For $E_0\in(0,1)$, the only minimum is $z^\star=\frac{1}{2}$ (disordered state). For $E_0>1$, the saddle point is given by
\begin{equation}
    E_0=\frac{1}{2(1-2z^\star)}\ln\frac{1-z^\star}{z^\star}, \quad \qty(E_0>1),
\end{equation}
which has two solutions $z^\star$ and $(1-z^\star)$ corresponding to the two ordered states resulting from spontaneous symmetry breaking.
These are the saddle point conditions in the main text (Eq.~4).

\subsubsection{Solution of $p_1$}
Let $p_1 = p_0\cdot\phi(x,y)$. By taking the hopping operator $\mathcal{L}_1$ to the continuum limit and substituting  the solution of $Q_0$ back to Eq.~\ref{Eq: p1 operator equation}, we arrive at the equation for $\phi$:
\begin{equation}
    \begin{aligned}
        \mathcal{L}_1 \qty(Q_0p_1)
         & =4Q_0p_0\qty(-x\pdv{\phi}{x}-y\pdv{\phi}{y}+ \frac{1}{Nz}\pdv[2]{\phi}{x}+\frac{1}{N(1-z)}\pdv[2]{\phi}{y})
        = -\mathcal{L}_2(Q_0p_0).
    \end{aligned}
\end{equation}
To obtain the right hand side, we expand Eq.~\ref{Eq: L2p0Q definition} to the next order in $N$. For example, the correction to $\lambda_{1,2}$ is
\begin{align}
    \lambda_1 & = \frac{a_1-b_1}{a_1+b_1} = 2z-1 + 2z(1-z)(x-y) + O\qty(N^{-1}),          \\
    \lambda_2 & = \frac{2a_0-N-a_1+b_1}{N-a_1-b_1} = 2z-1 - 2z(1-z)(x-y) + O\qty(N^{-1}).
\end{align}
Therefore,
\begin{align}
    \mathcal{L}_2(Q_0p_0) & = p_0Q_0 E_0\qty[2z(1-z)\psi_1 \qty(N(x-y)^2-\expval{N(x-y)^2}_0)] = 2E_0 p_0Q_0 \psi_1 \qty(z(1-z)N(x-y)^2-1),
\end{align}
where $\psi_1 = z e^{E_0(1-2z)} +(1-z)e^{-E_0(1-2z)}$, and the constant term is determined by $\sum_{a_1,b_1}\mathcal{L}_2(Q_0p_0)=\expval{\mathcal{L}_2(Q_0p_0)}_0=0$ which was the definition for $Q_0$.
Therefore, the equation for $\phi$ is
\begin{align}
    -x\pdv{\phi}{x}-y\pdv{\phi}{y}+ \frac{1}{Nz}\pdv[2]{\phi}{x}+\frac{1}{N(1-z)}\pdv[2]{\phi}{y}
     & = -\frac{E_0\psi_1}{2}\qty(z(1-z)N(x-y)^2-1),
\end{align}
The solution (for generic $z$) is:
\begin{align}
    \phi(x,y)=\frac{E_0}{4}\psi_1\qty(z(1-z)N\qty(x-y)^2-1).
\end{align}
Therefore, the probability distribution is
\begin{equation}
    {P = Q_0p_0\qty(1+\frac{\omega}{D}\phi) = Q_0p_0\qty(1+\frac{\omega E_0}{4D}\psi_1\qty(z(1-z)N\qty(x-y)^2-1))} + O\qty(\frac{\omega}{D})^2.
    \label{Eq:P up to p1 order}
\end{equation}

\subsubsection{Calculation of the alignment dissipation $\dot{w}_a$}
\begin{figure}
    \includegraphics[width=0.4\linewidth]{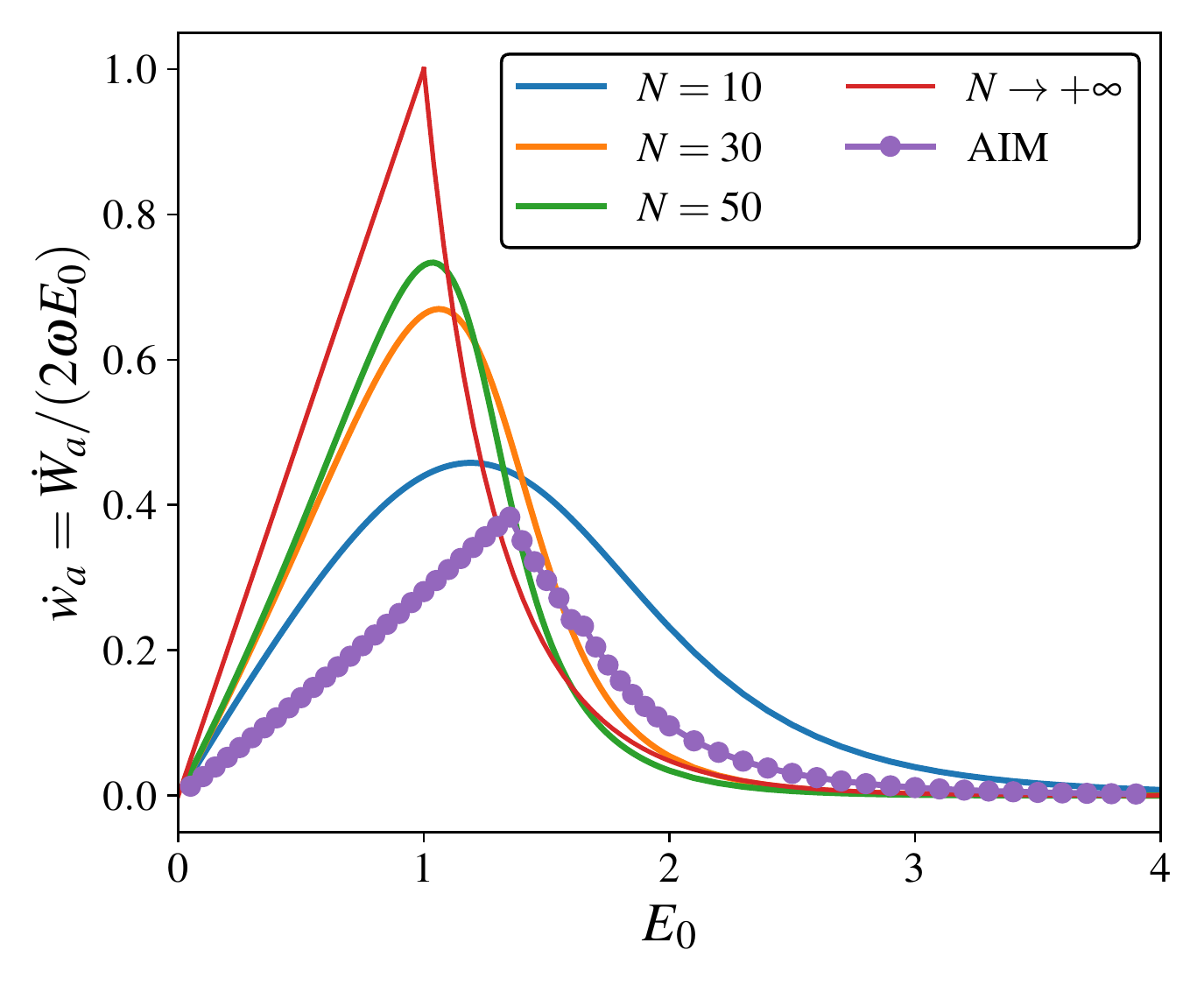}
    \caption{
        {Direct comparison between the alignment dissipation in the two-site model (solid lines) and the alignment dissipation per site in the full AIM (purple dots). The parameters are the same as Fig.~2B in the main text.}
    }
    \label{Fig: two-site raw}
\end{figure}

Now we compute the alignment dissipation $\dot{w}_a$ based on the probability distribution Eq.~\ref{Eq:P up to p1 order}.
\begin{align}
    \dot{w}_a = \frac{\dot{W}_a}{2\omega E_0} = \frac{1}{2\omega E_0} \sum
    \qty(J_+-J_-)\ln\frac{k_+}{k_-} = \frac{1}{2\omega E_0} \sum
    J_+\qty(1-\frac{J_-}{J_+})\ln\frac{k_+}{k_-}.
\end{align}
Due to symmetry between the two sites, we only need to consider flipping on site 1.
The rate ratio is
\begin{align}
    \ln \frac{k_+}{k_-} & = \ln \frac{\omega b_1 \exp\qty(E_0\frac{a_1-b_1}{a_1+b_1})}{\omega (a_1+1)\exp\qty(-E_0\frac{a_1-b_1+2}{a_1+b_1})} 
    =\ln\frac{1-z}{z} + 2E_0(2z-1) + 4E_0\frac{z(1-z)(x-y)}{1+xz+y(1-z)}.
\end{align}
The flux ratio is
\begin{align}
    \ln\frac{J_+}{J_-}
    = & \ln \frac{k_+}{k_-} + \ln\frac{P(a_0,a_1,b_1)}{P(a_0+1,a_1+1,b_1-1)}                        \\
    = & \qty(\ln\frac{1-z}{z} + 2E_0(2z-1) + 4E_0\frac{z(1-z)(x-y)}{1+xz+y(1-z)})+\alpha(z-z^\star)
    -\frac{\omega}{D}\frac{E_0\psi_1}{2} (x-y).
\end{align}
The alignment dissipation is
\begin{align}
    \dot{w}_a
    = & \frac{1}{2\omega E_0} \sum
    J_+\qty(1-\frac{J_-}{J_+})\ln\frac{k_+}{k_-}
    = \frac{1}{2\omega E_0}\cdot 2 \cdot\expval{
        b_1 \omega e^{E_0 m_1/\rho_1} \qty(1-\frac{J_-}{J_+})\ln\frac{k_+}{k_-}
    }                               \\
    = & \frac{1}{2E_0}\cdot\expval{
        N(1-z)(1+y) e^{E_0(2z-1)} \qty(1-\frac{J_-}{J_+})\ln\frac{k_+}{k_-}
    } = \expval{\sigma}.
\end{align}
$\sigma$ is the local alignment dissipation rate whose average over $P$ gives the steady-state $\dot{w}_a$. Importantly, it vanishes at the $(x,y,z)=(0,0,z^\star)$ (saddle point).
To calculate its expectation value, we need to expand $\sigma$ to the second order in $x$ and $y$, which is equivalent to the expansion in $a_1$ and $b_1$ shown in the main text. This demonstrates that the alignment dissipation emerges from the fluctuation of particle numbers.
Contributions from higher order terms are irrelevant in the $N\to\infty$ limit. The results are
\begin{align}
    \ln \frac{k_+}{k_-}     & = 4E_0\frac{z(1-z)(x-y)}{1+xz+y(1-z)}=4E_0z(1-z)(x-y),        \\
    \qty(1-\frac{J_-}{J_+}) & =\ln\frac{J_+}{J_-} = E_0\qty(4z(1-z)-\frac{\psi_1}{2})(x-y).
\end{align}
The alignment dissipation is
\begin{align}
    \dot{w}_a
    = & 2E_0 z(1-z)^2e^{E_0(2z-1)}\qty(4z(1-z)-\frac{\omega}{D}\frac{\psi_1}{2})\expval{N(x-y)^2}.
\end{align}
The last factor $(x-y)^2$ is averaged with respect to $p=p_0\qty(1+\frac{\omega}{D}\phi)$:
\begin{align}
    \expval{N(x-y)^2}
    = & \expval{N(x-y)^2}_0 + \frac{\omega}{D}\expval{\phi \cdot N(x-y)^2}_0
    =\frac{1}{z(1-z)}\qty(1+\frac{\omega E_0}{2D}\psi_1),
\end{align}
where the subscripts $0$ indicates averaging over $p_0$.
Therefore,
\begin{align}
    \dot{w}_a
    = & 2E_0 (1-z)e^{E_0(2z-1)}\qty(4z(1-z)-\frac{\omega}{D}\frac{\psi_1}{2})\qty(1+\frac{\omega E_0}{2D}\psi_1).
\end{align}
For the gas phase ($E_0\in(0,1)$), we have $z=\frac{1}{2}$ and $\psi_1 = 1$. The alignment dissipation is
\begin{align}
    \dot{w}_a = E_0\qty(1-\frac{\omega}{2D})\qty(1+\frac{\omega E_0}{2D}) = E_0\qty(1+\frac{\omega}{2D}(E_0-1)+O\qty(\frac{\omega}{D})^2),\quad E_0<1.
\end{align}
For the liquid phase ($E_0>1$), we have $z=z^\star(\neq\frac{1}{2})$ satisfying $E_0=\frac{1}{2(1-2z^\star)}\ln\frac{1-z^\star}{z^\star}$ and $\psi_1  =  \psi_1(z^\star)=2\sqrt{z^\star(1-z^\star)}$. The alignment dissipation is
\begin{align}
    \dot{w}_a
    = & E_0 \psi_1^3\qty(1-\frac{\omega}{2D}\psi_1^{-1})\qty(1+\frac{\omega E_0}{2D}\psi_1)=8E_0(z^\star)^{3/2}(1-z^\star)^{3/2}\qty(1+\frac{\omega}{2D}\qty(2E_0\sqrt{z^\star(1-z^\star)}-\frac{1}{2\sqrt{z^\star(1-z^\star)}})+O\qty(\frac{\omega}{D})^2)
\end{align}
To summarize:
\begin{equation}
    {\dot{w}_a = \begin{cases}
            E_0\qty(1+\frac{\omega}{2D}(E_0-1)+O\qty(\frac{\omega}{D})^2),\quad                                                                                                   & E_0<1. \\
            8E_0(z^\star)^{3/2}(1-z^\star)^{3/2}\qty(1+\frac{\omega}{2D}\qty(2E_0\sqrt{z^\star(1-z^\star)}-\frac{1}{2\sqrt{z^\star(1-z^\star)}})+O\qty(\frac{\omega}{D})^2),\quad & E_0>1.
        \end{cases}}
    \label{Eq: wa two-site perturbation result}
\end{equation}
The leading order terms are the dissipation presented in the main text, which, as demonstrated in Fig.~2B, is in good agreement with results in the full AIM after normalization.
The raw comparison prior to normalization is shown in Fig.~\ref{Fig: two-site raw}.

\begin{figure}[h!]
    \centering
    \includegraphics[width=0.8\textwidth]{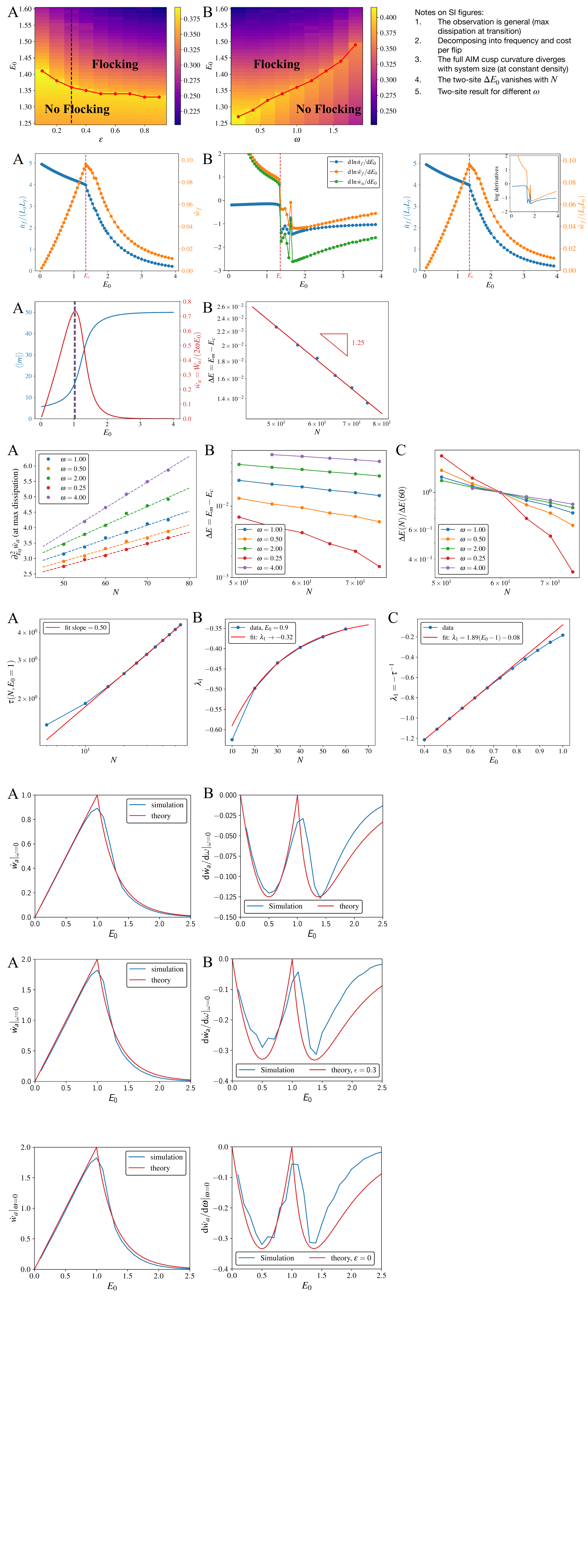}
    \caption{
        The alignment dissipation in the two-site model.
        Numerical results are obtained by solving the master equation for finite $N$ and extrapolating to infinite $N$. Analytical results are from Eq.~\ref{Eq: wa two-site perturbation result}.
        A: zeroth-order result at the fast diffusion limit (i.e. $\eval{\dot{w}_a}_{\omega=0}$);
        B: first-order correction (i.e. $\eval{\dv{\dot{w}_a}{\omega}}_{\omega=0}$ with $D=1$).
    }
    \label{Fig:TwoSite intercept slope}
\end{figure}

Both the leading order term and the $O\qty(\frac{\omega}{D})$ correction are in good agreement with numerical results obtained from numerically solving the master equation of the two-site model, which is shown in  Fig.~\ref{Fig:TwoSite intercept slope}.

\subsection{The energy-speed-sensitivity trade-off}

\subsubsection{Methods}
In the presence of the external field $h$, each spin contributes $\Delta H=-hs$ to the Hamiltonian of the local spin system. The flipping rates become
\begin{equation}
    k_{s\to(-s)} = \omega \exp\qty[
        -s\qty(E_0 \frac{m}{\rho}+h)
    ],
\end{equation}
which obey detailed balance within the local spin system. The probability distribution can be obtained by solving the master equation (with the modified flipping rates), and the dissipation can be calculated using the standard method.

The susceptibility $\chi$ follows its usual definition $\chi = \eval{\pdv{\expval{m}}{h}}_{h=0}$ where $m$ is the total magnetization.

As mentioned before, numerically determining $\expval{m}$ is difficult due to the absence of spontaneous symmetry breaking at finite $N$. To circumvent this, we need to manually break the symmetry by only considering the positive magnetization, namely $\expval{m}_+ = \frac{\sum_{m>0} m P(m)}{\sum_{m>0} P(m)}$. This calculation provides a good approximation deep in the flocking phase where $\abs{m}$ is large.

\subsubsection{Analytical results in the large $N$ limit}
In the large $N$ limit, the free energy landscape of the two-site model reads:
\begin{align}
    \frac{F(z)}{N} & = -\frac{\ln Q(z)}{N} = -\qty[g(z)+2E_0 z(1-z)] +2hz + O(N^{-1}),
\end{align}
where $g(z)=z\ln{z}+(1-z)\ln(1-z)$. The saddle point is determined by
\begin{align}
    \eval{\pdv{F}{z}}_{z=z^\star}=0\Rightarrow \ln\frac{z^\star}{1-z^\star}+2E_0(1-2z^\star)=2h.
\end{align}
The susceptibility $\chi=\pdv{m}{h}=2N\pdv{z^\star}{h}$ is given by:
\begin{align}
    2\frac{2N}{\chi} = \frac{1}{z^\star(1-z^\star)}-4E_0
    \Rightarrow
    \chi = \frac{N}{\frac{1}{4z^\star(1-z^\star)}-E_0}.
\end{align}

For a given flocking velocity $v=2D\epsilon \expval{s} = 2D\epsilon(2z^\star-1)$, we have
\begin{align}
    z^\star & = \frac{1}{2} + \frac{v}{4D\epsilon}                                                                                                                                                             \\
    E_0     & = \frac{1}{2(2z^\star-1)}\ln\frac{z^\star}{1-z^\star} = \frac{D\epsilon}{v}\ln\frac{1+\frac{v}{2D\epsilon}}{1-\frac{v}{2D\epsilon}}= \frac{\tanh^{-1}\frac{v}{2D\epsilon}}{\frac{v}{2D\epsilon}}
\end{align}
Therefore, the sensitivity $\chi$ and total dissipation rate $\dot{W}_\mathrm{tot}$ are
\begin{align}
    \frac{\chi}{N}                 & = \frac{1}{\frac{1}{4z^\star(1-z^\star)}-E_0}
    =\qty(\frac{1}{1-\qty(\frac{v}{2D\epsilon})^2}-\frac{\tanh^{-1}\frac{v}{2D\epsilon}}{\frac{v}{2D\epsilon}})^{-1} = f\qty(\frac{v}{2D\epsilon}),     \\
    \frac{\dot{W}_\mathrm{tot}}{N} & = 2D\epsilon  \ln\frac{1+\epsilon}{1-\epsilon} +\frac{2\omega E_0}{N} \cdot 8E_0 \qty[z^\star(1-z^\star)]^{3/2}    \\
                                   & =  2D\epsilon  \ln\frac{1+\epsilon}{1-\epsilon}+\frac{2\omega E_0^2}{N} \qty(1-\qty(\frac{v}{2D\epsilon})^2)^{3/2}
    =  2D\epsilon \ln\frac{1+\epsilon}{1-\epsilon} + O(N^{-1}).
\end{align}
The function $f(x) = \qty(\frac{1}{1-x^2}-\frac{\tanh^{-1}x}{x})^{-1}$ is monotonically decreasing in $x\in(0,1)$ and vanishes in the limit $\lim\limits_{x\to 1^-}f(x)=0$. The total dissipation is dominated by the motion part and increases monotonically with $\abs{\epsilon}$.

Using these expressions, we can examine the energy-speed-sensitivity trade-off:
\begin{itemize}
    \item For a given velocity $v$, $E_0$ decreases with $\epsilon$.
          The dissipation is minimized at $\epsilon=\frac{v}{2D}$ and $E_0\to\infty$. In this limit, the sensitivity vanishes, i.e. $\chi = Nf(1^-)=0$. The sensitivity is maximized in the limit $\epsilon \to 1$, which gives $\chi_\mathrm{max} = N f\qty(\frac{v}{2D})$. In this limit, however, the dissipation diverges.
    \item For $\epsilon \in \qty(\frac{v}{2D},1)$ and fixed $v$, both the sensitivity and the dissipation increases with $\epsilon$. For a fixed dissipation (therefore fixed $\epsilon$), the sensitivity decreases with the flocking velocity $v$. These relations lead to an energy-speed-sensitivity trade-off preventing the simultaneous improvement of all three properties.
    \item The energy-speed-sensitivity trade-off can be summarized with the following expression:
          \begin{align}
              \frac{\dot{W}_\mathrm{tot}}{2DN} = \mathcal{F}\qty[\frac{v}{2D}\mathcal{G}^{-1}\qty(\frac{\chi}{N})],
              \label{Eq:trade-off general}
          \end{align}
          where $\mathcal{F}(\epsilon) = \epsilon \ln\frac{1+\epsilon}{1-\epsilon}$ is a monotonically increasing function, and $\mathcal{G}^{-1}$ is the inverse of the following function
          \begin{align}
              \mathcal{G}(y) = \qty(\frac{y^2}{y^2-1}-\frac{y}{2}\ln\frac{y+1}{y-1})^{-1}, \quad (y>1)
          \end{align} which increases monotonically with $y$.
          Therefore, $\mathcal{G}^{-1}$ increases monotonically with $\chi$. Therefore, Eq.~\ref{Eq:trade-off general} clearly indicates that the dissipation needs to be increased to increase either speed or sensitivity, and that with a fixed dissipation increasing speed will decrease sensitivity.
    \item In the limit of strong coupling ($E_0\to\infty$), the relation between energy $\dot{W}_\mathrm{tot}$, speed $v$, and sensitivity $\chi$ can be obtained analytically. In this limit, the energy-speed-sensitivity trade-off is given by
          \begin{align}
              \frac{\dot{W}_\mathrm{tot}}{2DN} = \mathcal{F}\qty[\frac{v}{2D}\qty(1+\frac{\chi}{2N})],
              \label{Eq:trade-off limit1}
          \end{align}
          where $F(\epsilon) = \epsilon \ln\frac{1+\epsilon}{1-\epsilon}$ is a monotonically increasing function of $\epsilon$.
\end{itemize}

\section{The three-site active Ising model}
In this section, we present the the analytical solution to the three-site model, which not only captures the cusped dissipation maximum near the flocking transition but also explicitly show dependence on the bias $\epsilon$.

The state variables in the three-site model are:
\begin{itemize}
    \item $a_0$, $b_0$: total number of spins up/down;
    \item $a_{1,2,3}$, $b_{1,2,3}$: number of spins up/down on site 1, 2, and 3.
    \item $x_{1,2,3}=3a_{1,2,3}/a_0-1$, $y_{1,2,3}=3b_{1,2,3}/b_0-1$, $z=a_0/N$: continuous version of the state variables. We have $\sum_{i=1}^3x_i=\sum_{i=1}^3y_i=0$ by definition.
\end{itemize}
The perturbed solution is $P = Q_0\qty(p_0+\frac{\omega}{D}p_1)=Q_0p_0\qty(1+\frac{\omega}{D}\phi)$, where $p_0$ is the solution to the hopping operator, given by
\begin{equation}
    p_0 = \binom{a_0}{a_1}\binom{a_0-a_1}{a_2}\binom{b_0}{b_1}\binom{b_0-b_1}{b_2}\propto
    e^{-a_0h(x_1,x_2)-b_0h(y_1,y_2)},
\end{equation}
where
\begin{align}
    h(x_1,x_2) = \sum_i \frac{1+x_i}{3} \ln \frac{1+x_i}{3} = -\ln{3} + \frac{1}{3}\qty(x_1^2+x_1x_2+x_2^2) +O(x_1^4, x_2^4).
\end{align}
The perturbation theory follows the same approach albeit with two more state variables.
By following the same approach, it can be shown that the solution for $Q_0$ is identical to that in the two-site model:
\begin{align}
    Q_0(z) & \propto \exp\qty{-N\qty[2E_0z(1-z)+z\ln{z}+(1-z)\ln(1-z)] +O(1)}.
    \label{Eq: Q0 three-site}
\end{align}

It is useful to diagonalize the exponent of $p_0$ by a change of variable
\begin{equation}
    x_1 = \frac{u_1-u_2}{\sqrt{2}}, \quad x_2 = \frac{u_1+u_2}{\sqrt{2}}
    \Rightarrow
    h = -\ln 3 + \frac{1}{3}\qty(u_1^2+\frac{u_1^2-u_2^2}{2}+u_2^2) +O(u_1^4, u_2^4)=-\ln 3 +\frac{1}{2}u_1^2 + \frac{1}{6}u_2^2 +O(u_1^4, u_2^4).
\end{equation}
Therefore:
\begin{align}
    \ln p_0 = -\frac{Nz}{2}\qty(u_1^2+\frac{u_2^2}{3}) -\frac{N(1-z)}{2}\qty(v_1^2+\frac{v_2^2}{3}) + h.o.t.
\end{align}
leading to $\expval{u_1^2}=\frac{1}{Nz}$, $\expval{u_2^2}=\frac{3}{Nz}$, $\expval{v_1^2}=\frac{1}{N(1-z)}$, $\expval{v_2^2}=\frac{3}{N(1-z)}$.

\subsubsection{Solution of $p_1$}
Similar to the case in the two-site model, $p_1$ is determined by
\begin{equation}
    Q_0 \mathcal{L}_1 p_1
    = -\mathcal{L}_2(Q_0 p_0)
\end{equation}
where $\mathcal{L}_1$ is the hopping operator and $\mathcal{L}_2$ is the flipping operator.
However, the hopping operator now has explicit dependence on $\epsilon$:
\begin{align}
    \frac{\mathcal{L}_1 p_1}{ p_0}
    = & (1+\epsilon)\qty[a_2\phi(a_1+1,a_2-1,a_3)+a_3\phi(a_1,a_2+1,a_3-1)+a_1\phi(a_1-1,a_2,a_3+1)]                           \\
      & +(1-\epsilon)\qty[a_1\phi(a_1-1,a_2+1,a_3)+a_2\phi(a_1,a_2-1,a_3+1)+a_3\phi(a_1+1,a_2,a_3-1)]-N\phi(a_1,a_2,a_3) + \cc
\end{align}
where $\cc$ denotes the (conjugate) terms for spin down.
This means the same terms where $a,x,u$ are replaced by $b,y,v$, and $\epsilon$ should be replaced by $-\epsilon$.
The result in the continuum limit is
\begin{align}
    \frac{\mathcal{L}_1 p_1}{p_0}
    =       & -(3u_1-\epsilon u_2)\pu{1}-3(u_2+\epsilon u_1)\pu{2}                                                                                                                  \\
            & +\frac{1}{4Nz}\qty(3\qty(4-\sqrt{2}u_1-\sqrt{2}\epsilon u_1)\pu{1}^2+9\qty(4+\sqrt{2}u_1+\sqrt{2}u_2\epsilon)\pu{2}^2+6\sqrt{2}(u_2-3\epsilon u_1)\pu{1}\pu{2}) + \cc \\
    \approx & \qty[-(3u_1-\epsilon u_2)\pu{1}-3(u_2+\epsilon u_1)\pu{2}
        +\frac{1}{Nz}\qty(3\pu{1}^2+9\pu{2}^2)]\phi + \cc
\end{align}
The last line is obtained by dropping the $u_{1,2}$ terms in the quadratic coefficients.

Now we consider the flipping operator:
\begin{align}
    \frac{\mathcal{L}_2\qty(p_0Q)}{p_0}
    =  \sum_{i=1}^3 \left[ \frac{(a_0+1)Q(a_0+1)}{b_0}\qty(b_i e^{-E_0\qty(\lambda_i+\frac{2}{a_i+b_i})})+ \frac{(b_0+1)Q(a_0-1)}{a_0}\qty(a_ie^{E_0\qty(\lambda_i-\frac{2}{a_i+b_i})}) \right. \nonumber \\
        \left.- \qty(a_ie^{-E_0\lambda_i}+b_ie^{E_0\lambda_i})Q(a_0)\right],
\end{align}
where $\lambda_i=m_i/\rho_i$ is the average spin on site $i$. The leading order term [which is $O(N)$] vanishes with $Q=Q_0$ (given by Eq.~\ref{Eq: Q0 three-site}).
The expansion to the next order [which is $O(1)$] is:
\begin{align}
    \frac{\mathcal{L}_2\qty(p_0Q_0)}{p_0Q_0}
    = & \sum_{i=1}^3 \qty[ a_0\frac{1+y_i}{3} e^{-E_0\qty(\lambda_i+\frac{6}{N})}+ b_0 \frac{1+x_i}{3}e^{E_0\qty(\lambda_i-\frac{6}{N})} - \qty(a_0\frac{1+x_i}{3}e^{-E_0\lambda_i}+b_0\frac{1+y_i}{3}e^{E_0\lambda_i})] \\
    = & 2E_0 \psi_1\frac{z(1-z)}{3}\qty[\sum_{i=1}^3N(x_i-y_i)^2-\expval{\sum_{i=1}^3N(x_i-y_i)^2}_0]                                                                                                                    \\
    = & 2E_0 \psi_1\qty[z(1-z)N\qty(\qty(u_1-v_1)^2+\frac{1}{3}\qty(u_2-v_2)^2)-2],
\end{align}
where $\lambda_i = 2z-1 +2z(1-z)(x_i-y_i)$.
The constant term is determined by the equation for $Q_0$, namely $\sum_{a_1,b_1}\mathcal{L}_2(Q_0p_0)=\expval{\mathcal{L}_2(Q_0p_0)}_0=0$.  The resulting equation for $\phi$ is
\begin{align}
                & \frac{\mathcal{L}_1 p_1}{p_0} = -\frac{\mathcal{L}_2\qty(p_0Q_0)}{p_0Q_0} \\
    \Rightarrow & \qty[-(3u_1-\epsilon u_2)\pu{1}-3(u_2+\epsilon u_1)\pu{2}
        +\frac{1}{Nz}\qty(3\pu{1}^2+9\pu{2}^2) -(3v_1+\epsilon v_2)\pvv{1}-3(v_2-\epsilon v_1)\pvv{2}
        +\frac{1}{N(1-z)}\qty(3\pvv{1}^2+9\pvv{2}^2) ]\phi\nonumber                         \\
    =           & -2E_0\psi_1 \qty[z(1-z)N\qty((u_1-v_1)^2+\frac{1}{3}(u_2-v_2)^2)-2],
\end{align}
whose solution is
\begin{equation}
    {\phi = \frac{E_0\psi_1}{3}\qty(Nz(1-z)\qty[(u_1-v_1)^2+\frac{1}{3}(u_2-v_2)^2+\frac{2\epsilon}{3(3+\epsilon^2)}\qty(3\qty(u_1v_2-u_2v_1)+\epsilon(3u_1v_1+u_2v_2))] -2)},
\end{equation}
which gives the perturbed probability distribution by $P=Q_0p_0\qty(1+\frac{\omega}{D}\phi)$.

\subsubsection{Energy dissipation}
Now we calculate the alignment dissipation in the three-site model:
\begin{align}
    \dot{w}_a
    = & \frac{1}{2\omega E_0} \sum
    J_+\qty(1-\frac{J_-}{J_+})\ln\frac{k_+}{k_-}
    = \frac{1}{2\omega E_0}\cdot 3 \cdot\expval{
        b_1 \omega e^{E_0\lambda_1} \qty(1-\frac{J_-}{J_+})\ln\frac{k_+}{k_-}
    }                               \\
    = & \frac{1}{2E_0}\cdot\expval{
        N(1-z)(1+y_1) e^{E_0(2z-1)} \qty(1-\frac{J_-}{J_+})\ln\frac{k_+}{k_-}
    }.
\end{align}
The rate ratio follows from the two-site model:
\begin{align}
    \ln \frac{k_+}{k_-} & =4E_0z^\star(1-z^\star)(x_1-y_1),
\end{align}
where the higher order terms are omitted since they do not enter into the dissipation calculation.

\begin{figure}
    \centering
    \includegraphics[width=0.8\textwidth]{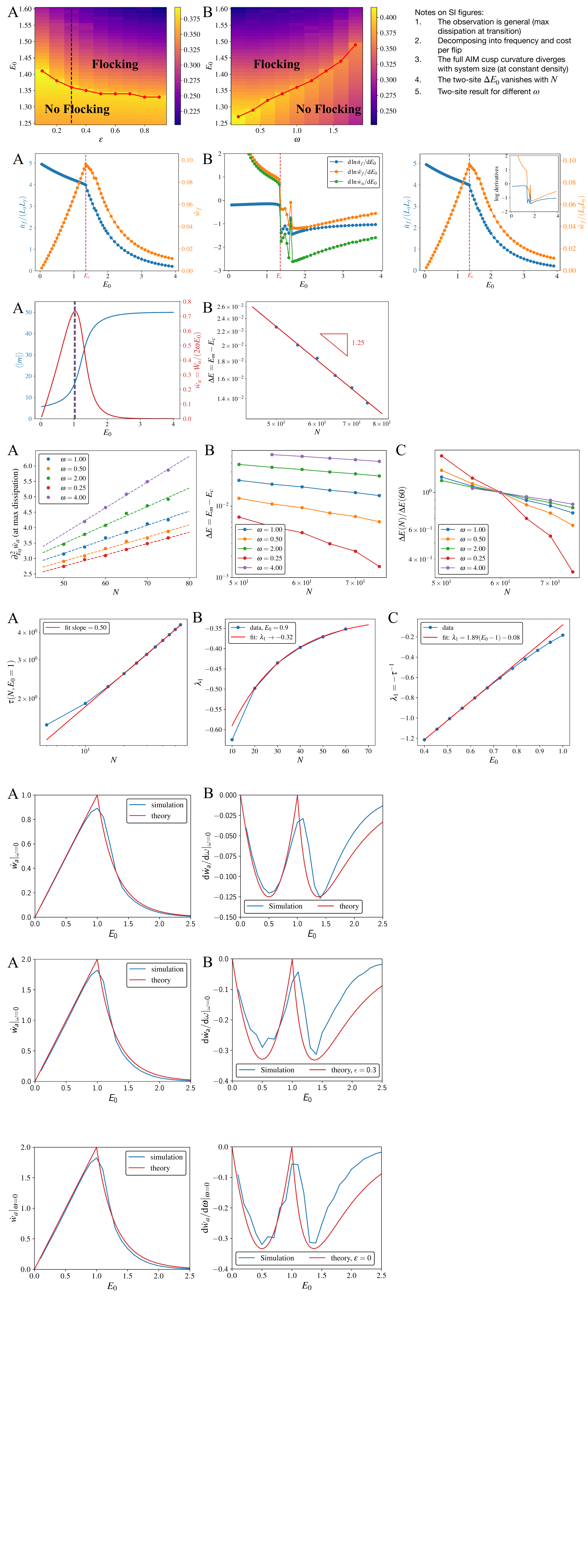}
    \caption{
        The alignment dissipation in the three-site model.
        Numerical results are obtained using the Gillespie algorithm for finite $N$ and extrapolating to infinite $N$. Analytical results are from Eq.~\ref{Eq: wa three-site perturbation result}.
        A: zeroth-order result at the fast diffusion limit (i.e. $\eval{\dot{w}_a}_{\omega=0}$);
        B: first-order correction (i.e. $\eval{\dv{\dot{w}_a}{\omega}}_{\omega=0}$ with $D=1$). $\epsilon=0.3$.
    }
    \label{Fig:ThreeSite intercept slope}
\end{figure}

The flux ratio is
\begin{align}
    \ln\frac{J_+}{J_-}
    = & \ln \frac{k_+}{k_-} + \ln\frac{P(a_0,a_1,b_1)}{P(a_0+1,a_1+1,b_1-1)}                                                                                                                       \\
    = & 4E_0z(1-z)(x_1-y_1) - \frac{4\omega }{9D} E_0\psi_1\qty[2(x_1-y_1)+(x_2-y_2) -\frac{\epsilon  (2 \epsilon  (x_1 z+y_1 (z-1))+x_2 z (\epsilon -3)+y2 (z-1) (\epsilon +3))}{\epsilon ^2+3}].
\end{align}
To the leading order, we have  $\qty(1-\frac{J_-}{J_+}) =\ln\frac{J_+}{J_-}$. Therefore, the alignment dissipation is
\begin{align}
    \dot{w}_a
    = & \frac{\sqrt{z(1-z)}}{2E_0}\expval{N\ln\frac{J_+}{J_-}\ln\frac{k_+}{k_-}} = 2 z^{3/2}(1-z)^{3/2}\expval{N(x_1-y_1)\ln\frac{J_+}{J_-}} \\
    = & 2 z^{3/2}(1-z)^{3/2}\qty(\expval{N(x_1-y_1)\ln\frac{J_+}{J_-}}_0+\frac{\omega}{D}\expval{\phi N(x_1-y_1)\ln\frac{J_+}{J_-}}_0).
\end{align}
The expectation values with respect to $p_0$ are calculated by substituting $(x,y)$ with $(u,v)$ for which the Hessian of $\ln p_0$ is diagonalized. The final result for the alignment dissipation is
\begin{equation}
    {\dot{w}_a = \begin{cases}
            2E_0\qty(1+\frac{\omega}{D}\frac{6+\epsilon^2}{3(3+\epsilon^2)}(E_0-1)+O\qty(\frac{\omega}{D})^2),\quad                                                                                                                           & E_0<1, \\
            16E_0(z^\star)^{3/2}(1-z^\star)^{3/2}\qty(1+\frac{\omega}{D} \frac{6+\epsilon^2(2-4z^\star(1-z^\star))}{3(3+\epsilon^2)}\qty(2E_0\sqrt{z^\star(1-z^\star)}-\frac{1}{2\sqrt{z^\star(1-z^\star)}})+O\qty(\frac{\omega}{D})^2),\quad & E_0>1.
        \end{cases}}
    \label{Eq: wa three-site perturbation result}
\end{equation}
The analytical solutions are in good agreement with numerical results (Fig.~\ref{Fig:ThreeSite intercept slope}), which evaluates both the zeroth order dissipation ($\frac{\omega}{D}\to 0$ ) and the first order correction [$O\qty(\frac{\omega}{D})$] using Gillespie simulations~\cite{gillespie_exact_1977}. Notably, the alignment dissipation Eq.~\ref{Eq: wa three-site perturbation result} has explicit dependence on $\epsilon$. Similar to the situation in the two-site model, the correction does not change the position and height of the cusp. It only modifies its slope on both sides.

\bibliography{flocking}